\documentclass[11pt,fleqn]{article}
\usepackage{setspace}
\usepackage{amsfonts}
\usepackage{amsmath}
\usepackage[margin=0.9in]{geometry}
\usepackage{enumerate}
\usepackage{eqnarray}
\usepackage{tabularx}
\usepackage{booktabs}
\usepackage{threeparttable}
\usepackage{graphicx}
\usepackage{mathabx}
\usepackage{natbib}
\usepackage[usenames,dvipsnames]{xcolor}
\usepackage{cite}
\usepackage{rotating}
\usepackage{fancyhdr}
\usepackage{soul}
\usepackage{titletoc}
\usepackage[toc,page]{appendix}
\newcommand{\sym}[1]{#1}
\usepackage{siunitx}
\usepackage{multirow}
\usepackage{lscape}
\usepackage{comment}
\usepackage{bbm}
\usepackage{cancel}
\usepackage{hyperref}
\hypersetup{
breaklinks=true,
    colorlinks,
    linkcolor={blue!30!black},
    citecolor={blue!30!black},
    urlcolor={blue!30!black}
}
\graphicspath{{./figures/}}
\usepackage{amssymb}
\newcommand*{\QEDB}{\hfill\ensuremath{\square}}%

\title{Correcting Attrition Bias using Changes-in-Changes$^*$}
\author{Dalia Ghanem\\ \small{UC Davis} \and Sarojini Hirshleifer\\ \small{UC Riverside} \and D\'esir\'e K\'edagni\\ \small{UNC-Chapel Hill}\\ \and Karen Ortiz-Becerra\\ \small{University of San Diego}}
\date{\today}

\begin{document}
\newtheorem{theorem}{Theorem}

\newtheorem{assumption}{Assumption}

\newtheorem{proposition}{Proposition}
\newtheorem{corollary}{Corollary}
\newtheorem{definition}{Definition}
\newtheorem{lemma}{Lemma}
\newtheorem{example}{Example}
\newtheorem{remark}{Remark}
\newenvironment{proof}[1][Proof]{\begin{trivlist}
\item[\hskip \labelsep {\bfseries #1}]}{\end{trivlist}}
\setcounter{page}{0}
\maketitle
\thispagestyle{empty}
\makeatletter
\def\blfootnote{\xdef\@thefnmark{}\@footnotetext}
\makeatother

\begin{abstract}
Attrition is a common and potentially important threat to internal validity in treatment effect studies.  We extend the changes-in-changes approach to identify the average treatment effect for respondents and the entire study population in the presence of attrition.  Our method, which exploits baseline outcome data, can be applied to randomized experiments as well as quasi-experimental difference-in-difference designs.  A formal comparison highlights that while widely used corrections typically impose restrictions on whether or how response depends on treatment, our proposed attrition correction exploits restrictions on the outcome model. We further show that the conditions required for our correction can accommodate a broad class of response models that depend on treatment in an arbitrary way. We illustrate the implementation of the proposed corrections in an application to a large-scale randomized experiment. 

\blfootnote{$^*$E-mail: \url{dghanem@ucdavis.edu}, \url{sarojini.hirshleifer@ucr.edu}, \url{dkedagni@unc.edu}, \url{kortizbecerra@sandiego.edu}. A Stata ado file to implement the proposed corrections is in progress.\\
We are grateful to the editor, associate editor, and two anonymous referees for constructive feedback that helped us improve the paper.}
\end{abstract}
\thispagestyle{empty}
\newpage
\section{Introduction}

Attrition is a common and potentially important source of selection bias in a range of treatment effect studies.  Attrition has long been recognized as a concern in settings that rely on panel data.\footnote{See, for example, \citet{FGM1998}, \citet{vdBL1998} and \citet{ZK1998}.}  In addition, as randomized experiments become a widely implemented methodology in applied economics, attrition tests and corrections are increasingly relevant to empirical practice \citep{MM2017,GHO2022}. The current empirical literature relies on a wide range of approaches to correct for attrition bias.  None of them, however, are specifically tailored to take advantage of the panel data available in many randomized experiments with baseline (pre-treatment) outcome data as well as in quasi-experimental difference-in-difference designs.

We propose a novel attrition correction based on the changes-in-changes (CiC) approach \citep{AI2006}. The correction is suitable for treatment effect settings where baseline outcome data are available. Extending the CiC framework to correct for attrition bias requires that the outcome is monotonic in a scalar unobservable and that the distribution of this unobservable conditional on treatment and response status is stable over time. While these assumptions are restrictive, they still allow the distribution of the outcome to vary across time, since that outcome can be a time-varying function of the unobservable.

The proposed method relies on a key insight: Under the extended CiC conditions, there are two transformations that relate the baseline outcome distribution to the distributions of the treated and untreated potential outcomes in the post-treatment period (Lemma \ref{lemma:CiC}). These two transformations are identical for all treatment-response subpopulations and can be identified using the treatment and control respondents. Using these transformations, we can identify not only the counterfactual distribution for the treatment and control respondents, but also the distribution of both potential outcomes for the treatment and control attritors. The identification of the average treatment effects for the respondents (ATE-R) as well as the entire study population (ATE) then follows immediately.  For outcomes that are continuous and strictly monotonic in the unobservable, the parameters of interest are point-identified.  For discrete outcomes, bounds can still be obtained for these objects under weak monotonicity restrictions (see Section \ref{app:discrete_extension} of the online appendix).

Since the CiC assumptions do not require random assignment, our approach is not only suitable for randomized experiments with baseline outcome data, but also quasi-experimental difference-in-difference designs. An advantage of random assignment, however, is that the CiC assumptions have an intuitive testable implication (without additional pre-treatment periods), which consists of the equality of the ``CiC-extrapolated'' average treatment effect on the treated and the untreated.\footnote{Without random assignment, the CiC assumptions can be tested in the presence of additional pre-treatment periods.}

We formally compare the assumptions required for CiC with those required for the inverse probability weighting approach (IPW) as well as widely used bounding approaches such as \citet{L2007} and \citet{BCGB2015}.\footnote{It is important to point out that, unlike CiC, all of these approaches require (conditional) random assignment of the treatment.}  In particular, the CiC assumptions can accommodate response models that depend on treatment without requiring monotonicity restrictions. By contrast, we show that the IPW assumptions do not allow response to depend on treatment status, whereas several bounding approaches such as \citet{L2007} and \citet{BCGB2015} require response to be monotonic in treatment status. The CiC identification approach instead exploits structural restrictions on the outcome model, whereas IPW and the aforementioned bounding approaches do not.  We further note the particular challenges of applying approaches aside from the CiC when the object of interest is the ATE.  \citet{L2007} and other related approaches do not recover this object, while the IPW approach requires a conditional missing-at-random assumption. We then consider the practical implications of these assumptions for a range of reasons for attrition in order to demonstrate how practitioners can assess the suitability of the CiC assumptions in empirical settings.

Finally, we illustrate the CiC attrition correction and compare it to existing approaches in an empirical example revisiting the randomized evaluation of the \emph{Progresa} cash transfer program. We find the CiC-corrected estimate of the ATE is significantly different from both the uncorrected treatment effect estimate as well as the IPW-corrected estimates.  We use several methods to consider the plausibility of the assumptions underlying these two approaches.  First, a test for attrition bias applied to this outcome finds that internal validity for the population is violated.\footnote{We implement the test of attrition bias proposed in \citet{GHO2022}.}  Second, we do not find evidence against the CiC identifying assumptions relying on their testable implication under random assignment (Remark \ref{rem:test}). Third, we conduct an analysis of correlates of attrition to consider plausible drivers of nonresponse and its implications for the corrections in this empirical example. 

This paper contributes to the literature on attrition corrections in treatment effect models which build on seminal work on sample selection \citep{H1976,H1979}.  It provides a tractable, nonparametric approach that exploits the presence of baseline outcome data through restrictions on the outcome model.  The standard Heckman correction (Heckit) approach assumes a parametric model where the treatment effect is homogeneous across individuals and the joint distribution of the errors in the outcome and response models is normal. \citet{MM2017} survey the field experiment literature and find that the most widely-used methods include IPW and various bounding approaches.\footnote{\citet{MM2017} also propose a modified version of the inverse probability weighting approach that uses additional data from an intense tracking phase.}  The former approach does not restrict the outcome model, but requires unconfoundedness and rules out the possibility that response depends on treatment status to identify the ATE-R. It further requires a conditional missing-at-random assumption to identify the ATE. There are several bounding approaches in the literature that do not impose any restrictions on the potential outcomes.  \citet{Manski1989} and \citet{HM1995,HM2000} provide bounds on treatment effect parameters that require minimal assumptions, however the bounds are typically wide in practice.  \citet{L2007} proposes bounds on the average treatment effect for the always-responders, a subset of the respondent subpopulation, assuming monotonicity restrictions on potential response. While our method imposes a monotonicity condition on the outcome variable, it does not impose monotonicity of response.   \citet{BCGB2015} show that tighter bounds on the average treatment effect for a subpopulation of respondents can be obtained by exploiting additional data, specifically the number of calls required to obtain a response. Their proposed bounds specifically exploit the monotonicity of the response in the number of maximal attempts to reach an individual to tighten \citeauthor{L2007}'s~(\citeyear{L2007}) bounds.\footnote{To do so, the authors assume that the maximum number of attempts to reach an individual is randomly assigned and excluded from the outcome. Relatedly, in the context of survey design, \citet{DiNardoetal2021} propose to randomly assign the probability of observation across survey participants.}

This paper proceeds as follows.  Section \ref{sec:metrics} introduces the model, discusses the implications of the time-invariance assumption for various response models, and provides the identification results with and without random assignment.  Section \ref{subsec:CIC_comps} conducts a formal comparison of the CiC correction with IPW and bounding corrections. Section \ref{sec:empirical}  applies the CiC and IPW corrections to the \textit{Progresa} randomized experiment. Section \ref{concl} concludes.

\section{Attrition Corrections via Changes-in-Changes \label{sec:metrics}}
\subsection{The Model and Parameters of Interest}\label{sec:model}
Let $Y_t$ and $D_t$ denote the observed outcome and treatment status in period $t$.  The treatment path is denoted by $D=(D_0,D_1)$.  To simplify notation, we denote the group membership by $G$, where $G=1$ for the treatment group which receives the treatment path $D=(0,1)$, and $G=0$ for the control group which receives the treatment path $D=(0,0)$, noting that $G=D_1$. We consider the following setting:   
\begin{eqnarray}\label{seq1}
\left\{ \begin{array}{lcl}
     Y_{t} &=& Y_0(0) \mathbbm{1}\{t=0\}+Y_1(G) \mathbbm{1}\{t=1\},  \\ 
     R &=& GR(1)+(1-G)R(0).
     \end{array} \right.
\end{eqnarray}
$Y_{0}(0)$ denotes the untreated potential outcome in the baseline period ($t=0$) and $Y_1(d)$ denotes the potential outcome for $d=0,1$ for the follow-up period ($t=1$). The outcome variable in the baseline period ($t=0$) is always observed, whereas the outcome variable in the follow-up period ($t=1$) is observed only if $R=1$. $R$ specifically denotes response in the follow-up period with $R(d)$ denoting the potential response given treatment status $d=0,1$. We assume there are no response issues in the baseline period $(t=0)$.

In this paper, we are interested in identifying the average treatment effects for the treated and untreated respondents (ATT-R and ATU-R), for the respondents (ATE-R), and for the study population\footnote{The study population is the population that the sample is drawn from.} (ATE), defined as follows:
 \begin{eqnarray*}
 \text{ATT-R} &=& E[Y_{1}(1)-Y_{1}(0)|G=1,R=1],\\
 \text{ATU-R} &=& E[Y_{1}(1)-Y_{1}(0)|G=0,R=1],\\
 \text{ATE-R} &=& E[Y_{1}(1)-Y_{1}(0)|R=1],\\
 \text{ATE} &=& E[Y_{1}(1)-Y_{1}(0)].
 \end{eqnarray*}
We obtain a random sample of the vector $(G, R, Y_{0}, Y_{1}^*)$, where all random variables are observed except $Y_1^*$ which is only observed when $R=1$.\footnote{In our setting, we observe $Y_0$ for all units. As a result, $R$ equals to one for units with observations of both $Y_0$ and $Y_1^*$, whereas it equals zero for units with observed $Y_0$ but missing $Y_1^*$. Thus, $R$ is a response-group indicator, whereas $G$ is a treatment-group indicator.} 

In order to identify the above parameters, we assume that the potential outcomes are given by the following model,
\begin{eqnarray*}Y_0(0)&=&\mu_0(0,U_0),\\
Y_1(d)&=&\mu_1(d,U_1)~\text{for $d=0,1$}.
\end{eqnarray*}
The variable $U_{t}$ denotes the unobserved heterogeneity in the outcome, which is assumed to be scalar. In the following, we present our identifying assumptions which impose specific restrictions on $\mu_t(\cdot)$ and $U_t$. Let $F_Y$ denote the cumulative distribution function of a random variable $Y$. 

\begin{assumption}\label{ass:Unobs} (Distribution of Unobservable) \begin{enumerate}[1.]
\item $U_{0}|G,R\overset{d}{=}U_{1}|G,R$.\label{TI} \item $F_{U_t|G,R}(u|g,1)$ is continuous and strictly increasing in $u$ for $g=0,1$.\label{continuity}\end{enumerate}
\end{assumption}
\begin{assumption}\label{ass:Mon} (Monotonicity of Structural Function)

\begin{enumerate}[1.]
\item $\mu_{t}(0,u)$ is strictly increasing in $u$ for $t=0,1$. \label{MonUPO}
\item  $\mu_{1}(1,u)$ is strictly increasing in $u$. \label{MonTPO}
\end{enumerate}
\end{assumption}

Assumption \ref{ass:Unobs}.\ref{TI} states that the distribution of unobservables that affect the outcome ($U_t$) is stable over time within each treatment-response subgroup. It is similar to Assumption 3.3 in \citet{AI2006}, except that we condition on the observed response status. This assumption rules out time variability in the distribution of unobservables within each treatment-response subpopulations, but it admits selection into the program and survey response as it allows for differences in the distribution of $U_t$ and $Y_t(d)$ across the four subgroups. For a more detailed discussion of the plausibility of this assumption considering both the unobservable determinants of the outcome and response, see Section \ref{sec:response}. We further impose Assumption \ref{ass:Unobs}.\ref{continuity} to ensure that the distribution function of $Y_t|G=g,R=1$ is invertible for $g=0,1$. 

Assumption \ref{ass:Mon}.\ref{MonUPO} implies that for each period, the untreated potential outcome is strictly increasing in the unobserved heterogeneity. It is the same as Assumption 3.2 in \citet{AI2006}. Assumption \ref{ass:Mon}.\ref{MonTPO} requires that the treated potential outcome is strictly increasing in the unobserved heterogeneity.\footnote{For instance, if $U_t$ is a single characteristic such as ability and the outcome is profits, Assumption \ref{ass:Mon} implies that higher levels of ability correspond to higher potential profits.} These two monotonicity assumptions are the main driver of our identification results as we show in Lemma \ref{lemma:CiC}.  Assumption \ref{ass:Mon} is automatically satisfied when the structural function is additively separable, such as $Y_t(d)=\gamma_t d+\lambda_t(1-d)+U_t$, where $U_t=\alpha+\varepsilon_t$ includes a time-invariant and time-varying component. It holds, however, for the broader class of potentially nonlinear, monotonic transformations, $Y_t(d)=\mu_t(d,U_t)$. We provide further examples in Section \ref{sec:response}.

The strict monotonicity conditions imposed in Assumption \ref{ass:Mon} have consequences for the interpretation of $U_t$ and the time-invariance condition in Assumption \ref{ass:Unobs}. These conditions specifically imply that $U_t$ may be viewed as the normalized outcome, specifically $U_t=\mu_t^{-1}(0;Y_t(0))$, where $\mu_t^{-1}(0;y)$ denotes the inverse of $\mu_t(0,u)$. In light of Assumption \ref{ass:Mon},  Assumption \ref{ass:Unobs} requires that the change in the potential outcome distribution across time is only driven by the change in the monotonic structural function, $\mu_t(d,\cdot)$.  This allows for the outcome distribution to change across time as long as it admits a normalization that renders its distribution stable across time.\footnote{To illustrate this point, consider a setting where the potential outcome is given by the location-scale model: $\mu_t(d,U_t)=\sigma_t^dU_t+\alpha_t^d$. Then, the conditional time-invariance assumption requires that the change in the potential outcome distribution across time is solely due to the change in $\mu_t(d,\cdot)$ as follows, $F_{Y_t(d)|G,R}(y)=P(\sigma_t^dU_t+\alpha_t^d\leq y|G,R)=P\left(U_t\leq \frac{y-\alpha_t^d}{\sigma_t^d}\mid G,R\right)=F_{U_0|G,R}\left(\frac{y-\alpha_t^d}{\sigma_t^d}\right)$.}

\begin{remark}[Discrete Outcomes]
 We extend our attrition corrections to discrete outcomes allowing for weak instead of strict monotonicity of the structural function $\mu_t(d,U_t)$ in Section \ref{app:discrete_extension} of the online appendix.\footnote{Theoretically, the strict monotonicity condition can hold in the discrete outcome case if the unobservable is also discrete.  However, this would rule out the most well-known models for limited dependent variables, such as linear-index models as discussed in \citet{AI2006}.}  Similar to \citet{AI2006}, we can only provide partial identification results for this case.
\end{remark}

Since we will provide attrition corrections in randomized experiments, we formally define the assumption of random assignment.

\begin{assumption}\label{ass:RA} (Random Assignment)
$(Y_{0}(0),Y_1(0),Y_{1}(1),R(0),R(1))\perp G$.
\end{assumption}
Assumption \ref{ass:RA} states that the individuals are randomly assigned to the treatment $(G=1)$ and control $(G=0)$ groups. This assumption applies to randomized experiments with simple and cluster randomization designs.  Below we provide identification results with and without random assignment.

In Section \ref{sec:id}, we rely on the strict monotonicity of the structural function (Assumption \ref{ass:Mon}) together with Assumption \ref{ass:Unobs} to provide point-identification results for continuously distributed random variables with strictly increasing distributions. Before we do so, we examine the conditions that are necessary and sufficient for a given response model to satisfy the conditional time-invariance assumption.

\subsubsection{Conditional Time Invariance and Response Models\label{sec:response}} 
In this section, we provide necessary and sufficient conditions for the conditional time-invariance assumption (Assumption \ref{ass:Unobs}.\ref{TI}) required for our identification result. These conditions are useful in applications where researchers may have \emph{a priori} information on the sources of non-response in their setting. To illustrate the conditional time-invariance assumption further, we discuss its plausibility in the context of examples of unobservable determinants of the outcome and response.

We first consider some examples of unobservable determinants of the outcome. In some settings, the conditional time-invariance assumption is natural. For instance, consider a setting in which the treatment is a microcredit program, the outcome of interest is profits, and $U_t$ is an unobserved determinant of profits. If the relevant unobserved heterogeneity is a trait that is not typically viewed as changing over time ($U_0=U_1$), such as ability or risk preferences, then the conditional time-invariance assumption is trivially satisfied. 

Alternatively, if profits are determined by a time-varying unobservable ($U_0\neq U_1$), the conditional time-invariance assumption can still be satisfied.  To determine this, however, it is crucial to relate the interpretation of $U_t$ to the structure imposed on the potential outcomes. To fix ideas, let $\tilde{U}_t^d=U_t\sigma_t^d+\alpha_t^d$ denote a health shock, which may have a different mean and variance across time as well as by treatment status. For instance, if the follow-up period coincides with a season during which malaria is endemic, then the health shock can have a negative mean and lower standard deviation to signify the higher likelihood of receiving a bad health shock, specifically malaria. That is, the mean and standard deviation of the health shock can vary by treatment status as well such that receiving the treatment can change the conditional distribution of health shocks faced by the individuals. More generally, if profits, given by $Y_t(d)=\tilde{\mu}_t(d,\tilde{U}_t^d)$, are a strictly monotonic transformation of $\tilde{U}_t^d$, then they are also a strictly monotonic transformation of $U_t$, the normalized health shock.\footnote{To see this, recall that $U_t=(\tilde{U}_t^d-\alpha_t^d)/\sigma_t^d$ and $\tilde{U}_t^d=\tilde{\mu}_t^{-1}(d;Y_t(d))$, so $\mu_t(d,U_t)=\tilde{\mu}_t(d,\tilde{U}_t^d)$ is a composition of two monotonic transformations. To normalize the outcome and obtain $U_t$, we invert the composition of the two functions, $U_t=\left(\tilde{\mu}_t^{-1}(d;Y_t(d))-\alpha_t^d\right)/\sigma_t^d$.} Thus, while the normalized health shock has to obey the conditional time-invariance assumption (Assumption \ref{ass:Unobs}.\ref{TI}), the conditional distribution of $\tilde{U}_t^d$ is allowed to change over time and by treatment status.

To further analyze the plausibility of this conditional time-invariance assumption, it is helpful to understand how it relates to the response model. Thus, we consider unobservable determinants of response and characterize the restrictions that are necessary and sufficient for the conditional time-invariance assumption.

\begin{proposition}\label{prop:CIC_selection}
Suppose that $0<P(R=r|G)<1$ for $r=0,1$. Suppose there exists a random vector $V$ such that $(U_0, V ) | G \overset{d}{=}(U_1, V ) | G$ and $R=\varphi(G,V)$, where $\varphi(\cdot)$ is a measurable function of $(V,G)$. Then, 
$U_{0}|G,R\overset{d}{=}U_{1}|G,R$ (Assumption \ref{ass:Unobs}.\ref{TI}). 

Conversely, suppose $U_{0}|G,R\overset{d}{=}U_{1}|G,R$. Then there exists a random vector $V$ and a measurable function $\varphi(.)$ such that $(U_0, V ) | G \overset{d}{=}(U_1, V ) | G$ and $R=\varphi(G,V)$.
\end{proposition}

All proofs are in the appendix. This proposition provides a condition on the unobservables determining response that holds iff the distribution of $U_t|G,R$ is time-invariant. The proposition specifically states that if response is determined by a vector of unobservables $V$ as well as treatment, then Assumption \ref{ass:Unobs}.\ref{TI} would hold iff the joint distribution of $(U_t,V)$ is time invariant conditional on $G$.\footnote{It is important to note that this condition allows for $U_t$ and $V$ to be dependent conditional on $G$. To see this, note that $F_{U_t,V|G}(u,v)=C_{U_t,V|G}(F_{U_t|G}(u),F_{V|G}(u))$ (Sklar theorem), where $C_{U_t,V \vert G}$ denotes the copula between $U_t$ and $V$ conditional on G. As a result, while the time invariance of the joint distribution requires the time invariance of the copula $C_{U_t,V|G}$ and the distribution of $U_t|G$, it does not restrict the type of copula that governs the dependence between $U_t$ and $V$.} Note that under random assignment, which is given by $(V,U_0,U_1)\perp G$ in our context, the condition in Proposition \ref{prop:CIC_selection} would be replaced by its unconditional version, $(U_0,V)\overset{d}{=}(U_1,V)$.

Next, we consider an example of a response model and discuss the condition in Proposition~\ref{prop:CIC_selection} in the context of this example. 
\begin{example}\label{ex:prop4}
Let $V$ denote some unobservable determinant of response, such as potential to migrate or reciprocity.  Response is a threshold-crossing model of $V$, where the threshold depends on the treatment status, such that $c_0$ ($c_0+c_1$) is the threshold for the control (treatment) group,
\begin{align}R&=1\{V\geq c_0+c_1G\}.\label{eq:R_example1}\end{align}
Note that if $c_1<0$, the threshold to respond in the treatment group would be lower than the control group, whereas if $c_1>0$, the threshold to respond is higher in the treatment than in the control group. 

For simplicity, assume that random assignment holds, $(V,U_0,U_1)\perp G$. Now suppose that $(V,U_0,U_1)\sim N(0,\Sigma)$, where 
$$\Sigma=\left(\begin{array}{ccc}1 & \delta  &  \rho \\ \delta & 1 &  \rho\\ \rho &  \rho & 1\end{array}\right).$$ By Proposition \ref{prop:CIC_selection}, it follows that the conditional time-invariance assumption required for the CiC attrition correction is satisfied in this example, since the joint distribution of $U_t$ and $V$ is time-invariant. As a result, the unobservable determinant of the potential outcomes and response can be dependent as long as the joint distribution of their unobservable determinants is stable over time.

Next, consider a setting where nonresponse is determined by whether individuals stay in their current location or migrate, and whether they are willing or reluctant to respond to a survey. Let $S=\xi(G,V_S)$ denote a binary variable that equals one if an individual stays in their current location and zero otherwise, whereas $W=\omega(G,V_W)$ equals one if an individual is willing to respond and zero otherwise. In this case, response status equals the product of $S$ and $W$, 
\begin{eqnarray}R=S\cdot W=\xi(G,V_S)\omega(G,V_W).\label{eq:response_multiple}\end{eqnarray}
In this setting, the determinants of response consist of the determinants of migration and the determinants of willingness to respond, $V=(V_S',V_W')$. If a researcher is willing to assume the joint distribution of $(V,U_t)$ is stable across time, then by Proposition \ref{prop:CIC_selection} the conditional time-invariance assumption required for the CiC correction holds, regardless of the functional form and properties of the response model.

\QEDB
\end{example}
The above example demonstrates that the conditions in Proposition \ref{prop:CIC_selection} do not impose restrictions on the functional form of $R$ and are consistent with a multidimensional $V$. This feature is especially attractive in settings where response is determined by multiple factors.

Proposition \ref{prop:CIC_selection} provides a general necessary and sufficient condition that allows for dependence between $V$, $U_0$ and $U_1$ and obeys time-invariance restrictions as illustrated in the above example. The proposition is not explicit, however, on what precise conditions would imply the time invariance of $(V,U_t)|G$ if $V$ is a function of $U_0$ and $U_1$. The following corollary addresses this issue by examining a special case of Proposition \ref{prop:CIC_selection} where response is determined by a function of $U_0$ and $U_1$.\footnote{It is worth noting that if response is solely determined by baseline outcome, $R=f(Y_0)$, random assignment ($(Y_0(0),Y_1(0),Y_1(1),R(0),R(1))\perp G$) implies  $(Y_1(0),Y_1(1))\perp G|R$, which would yield a case where no correction would be warranted and the ATE-R would be identified from the simple difference in means between treatment and control respondents. However, while this is a theoretically interesting case, it is not very relevant from a practical perspective since response at follow-up is also likely affected by unobservable factors in the follow-up period and the treatment status itself.} We emphasize that unlike Proposition \ref{prop:CIC_selection}, the condition in the following corollary is merely sufficient, and not necessary, for Assumption \ref{ass:Unobs}.\ref{TI} to hold.

\begin{corollary}\label{cor:CIC_selection}
Suppose that $0<P(R=r|G)<1$ for $r=0,1$. Suppose further that $R=\psi(G,U_0,U_1)$.\\
If the mapping $(u_0,u_1) \longmapsto \psi(.,u_0,u_1)$ is symmetric in its arguments $(u_0,u_1)$, and $F_{U_{0},U_1|G}(u_0,u_1)$ is exchangeable in $U_0$ and $U_1$, then $U_{0}|G,R\overset{d}{=}U_{1}|G,R$ (Assumption \ref{ass:Unobs}.\ref{TI}). 
\end{corollary}
The above corollary establishes that if response is determined by $G$, $U_0$, and $U_1$, then for the time-invariance condition to hold conditional on $G$ and $R$, it is sufficient for response to be symmetric in $U_0$ and $U_1$ and the distribution of $(U_0,U_1)$ conditional on $G$ to be exchangeable in $U_0$ and $U_1$.\footnote{It is worth noting that the exchangeability condition implies the time invariance of the distribution of $U_t$ conditional on $G$, specifically $U_0|G\overset{d}{=}U_1|G$.} The following example provides an example of a response model that obeys these conditions.

\begin{example}\label{ex:cor1}
Suppose that the outcome is earnings. Under the strict monotonicity condition in Assumption \ref{ass:Mon}, $U_t$ can be viewed as the normalized earnings, $U_t=\mu_t^{-1}(0,Y_t(0))$. The following model for response requires that once the sum of normalized earnings exceeds a particular threshold individuals might choose to migrate and therefore they will not respond to the survey,
\begin{align}R=1\{U_0+U_1\leq c_0+c_1G\}.\label{eq:R_ex1}\end{align}
Here, as in Example \ref{ex:prop4}, we can allow the threshold to depend on whether one is in the treatment or control group. This response model is consistent with the symmetry condition in Corollary \ref{cor:CIC_selection}. 
For the time-invariance condition in Assumption \ref{ass:Unobs}.\ref{TI} to hold, we would further require the distribution of $(U_0,U_1)$ to be exchangeable conditional on $G$ (e.g. if $(U_0,U_1)|G$ are jointly normal with the same mean and variance).

\QEDB
\end{example}

Finally, it is important to consider an example where response depends on the unobservable determinant of the follow-up outcome, $U_1$. This example neither obeys the conditions in Corollary \ref{cor:CIC_selection} nor Assumption \ref{ass:Unobs}.\ref{TI}.

\begin{example}\label{ex:cor2}
Consider the same setting in Example \ref{ex:cor1}. Suppose, however, that response depends on the normalized outcome in the follow-up period,
$$R=1\{U_1\leq c_0+c_1G\},$$
where $\left(\begin{array}{c}U_0\\ U_1\end{array}\right) \vert G\sim N(G.\iota,\rho I_2)$, $I_2$ is a $2 \times 2$ identity matrix, $\iota=\left(\begin{array}{c}1\\ 1\end{array}\right)$, and $0 < \rho < 1$. Since the vector $(U_0,U_1)'$ is Gaussian, we can write $U_0=\rho U_1 + \varepsilon$, where $\varepsilon \perp U_1$ and $\varepsilon \sim N(0,1-\rho^2).$ Therefore, $U_0 \vert G, R$ does not follow the same distribution as $U_1 \vert G, R$. Hence, Assumption \ref{ass:Unobs}.\ref{TI} fails to hold. 
\QEDB
\end{example}

\subsection{Identification Results}\label{sec:id}
In this section, we outline how the CiC identification approach can be applied to point-identify our objects of interest.  We provide results both for the respondent subpopulation and study population.  Let $\mathbb Y$ denote the support of the random variable $Y$, and $\mathbb{Y}_{g,r}^{d,t}$ denote the support of $Y_{t}(d)|G=g,R=r$. Define $F_Y^{-1}(q)=\inf\{y\in \mathbb Y \vert F_Y(y) \geq q\}$.

Before we proceed to our main identification results, the following lemma helps us understand how Assumptions \ref{ass:Unobs} and \ref{ass:Mon} can allow us to ``extrapolate'' not only to the respondent subpopulations but also to the attritor subpopulations. 
\begin{lemma}\label{lemma:CiC} Suppose that Assumptions \ref{ass:Unobs} and \ref{ass:Mon}.\ref{MonUPO} hold, then:
\begin{enumerate}[1.]\item \label{T0}For $g=0,1$, $r=0,1$,
\begin{align*}
   (i)&~ F_{Y_1(0)|G=g,R=r}(y)=F_{Y_0(0)|G=g,R=r}(T_0(y))~ for ~ y\in\mathbb{Y}_{g,r}^{0,1},\\
   (ii)&~ T_0(y)=F_{Y_0|G=0,R=1}^{-1}(F_{Y_1|G=0,R=1}(y)) \quad \quad ~for~y\in \mathbb{Y}_{0,1}^{0,1},
\end{align*}
where $T_0(y)=\mu_0(0,\mu_1^{-1}(0;y))$ and $\mu_1^{-1}(0;y)$ denotes the inverse of $\mu_1(0,u)$.
\item \label{T1} Suppose further that Assumption \ref{ass:Mon}.\ref{MonTPO} holds. For $g=0,1$, $r=0,1$,
\begin{align*}
   (i)&~ F_{Y_1(1)|G=g,R=r}(y)=F_{Y_0(0)|G=g,R=r}(T_1(y))~ for ~ y\in\mathbb{Y}_{g,r}^{1,1},\\
   (ii)&~      T_1(y)=F_{Y_0|G=1,R=1}^{-1}(F_{Y_1|G=1,R=1}(y))\quad \quad ~for ~ y\in \mathbb{Y}_{1,1}^{1,1},
\end{align*}
where $T_1(y)=\mu_0(0,\mu_1^{-1}(1;y))$ and $\mu_1^{-1}(1;y)$ denotes the inverse of $\mu_1(1,u)$.
\end{enumerate}
\end{lemma}
Lemma \ref{lemma:CiC}.\ref{T0}(i) shows that under the time-invariance assumption (Assumption \ref{ass:Unobs}.\ref{TI}) and the strict monotonicity of the untreated potential outcome (Assumption \ref{ass:Mon}.\ref{MonUPO}), the distribution of the untreated potential outcome for any treatment-response subpopulation in the follow-up period at a given $y$ equals the distribution of the untreated potential outcome of that subpopulation in the baseline period evaluated at $T_0(y)$, where the transformation, $T_0(\cdot)$, is the same for all treatment-response subpopulations.  Since we observe the distribution of the untreated potential outcome of the control respondents in both baseline and follow-up periods, Lemma \ref{lemma:CiC}.\ref{T0}(ii) shows that we can identify $T_0(y)$ for $y\in\mathbb{Y}_{0,1}^{0,1}$ using the control respondents by the continuity and strict monotonicity of the outcome distribution (Assumption \ref{ass:Unobs}.\ref{continuity}).  

If we also impose the strict monotonicity assumption on the treated potential outcome (Assumption \ref{ass:Mon}.\ref{MonTPO}), Lemma \ref{lemma:CiC}.\ref{T1}(i) shows that the treated potential outcome distribution for any treatment-response subpopulation at a given value $y$ equals the distribution of the untreated potential outcome of that subpopulation in the baseline period evaluated at $T_1(y)$, where the transformation, $T_1(\cdot)$, is the same for all treatment-response subpopulations.  Since we observe the untreated potential outcome in the baseline period and the treated potential outcome in the follow-up period for the treatment respondents, we can use them to identify $T_1(y)$ for $y\in\mathbb{Y}_{1,1}^{1,1}$ (Lemma \ref{lemma:CiC}.\ref{T1}(ii)). 

In sum, Lemma \ref{lemma:CiC} shows that we can use the control and treatment respondents to identify $T_0(y)$ and $T_1(y)$ on their respective support.  Since $T_0(y)$ and $T_1(y)$ are the same for all subpopulations, the identification of the distribution of an unobserved potential outcome for a given treatment-response subpopulation follows immediately assuming that we can observe the baseline outcome distribution for this subpopulation and that additional support conditions hold. We finally note that Lemma \ref{lemma:CiC} does not require random assignment.  This allows us to provide identification results for our parameters of interest without random assignment (Assumption \ref{ass:RA}).
\subsubsection{Identification Results for the Respondent Subpopulation}
In the following, we provide identification results for the average treatment effects for the respondent subpopulations. Note that since individuals choose to respond or not, treatment is no longer randomly assigned conditional on response without further restrictions. As a result, the results in this section do not require random assignment. They instead exploit the conditional time-invariance assumption as well as the structural assumptions imposed by the CiC conditions. The identification results in this section constitute a direct application of CiC to the respondent subpopulation.

We first establish the identification of the ATT-R, since it requires the strict monotonicity condition on the untreated potential outcome only (Assumption \ref{ass:Mon}.\ref{MonUPO}) in addition to the assumptions on the unobservables. Let $\mathbb{U}_{g,r}$ denote the support of $U_{0}|G=g, R=r$ for $g=0,1$, and $r=0,1$. For two sets $\mathbb{A}$ and $\mathbb{B}$, $\mathbb{A}\subseteq \mathbb{B}$ denotes that $\mathbb{A}$ is contained in $\mathbb{B}$.

\begin{proposition}[Identification of the ATT-R]\label{prop:TR} Suppose that Assumptions \ref{ass:Unobs} and \ref{ass:Mon}.\ref{MonUPO} hold. Suppose further that $\mathbb{U}_{1,1}\subseteq \mathbb{U}_{0,1}$. Then, 
\begin{align}&F_{Y_{1}(0)|G=1,R=1}(y)=F_{Y_{0}|G=1,R=1}(F_{Y_{0}|G=0,R=1}^{-1}(F_{Y_{1}|G=0,R=1}(y)))\quad for\quad y\in\mathbb{Y}_{1,1}^{0,1},\label{eq:TR_counterfactual}\\
&\text{ATT-{R}}=E[Y_{1}|G=1,R=1]-E[F_{Y_{1}|G=0,R=1}^{-1}(F_{Y_{0}|G=0,R=1}(Y_{0}))|G=1,R=1]. \label{eq:TR_ATE}
\end{align}
\end{proposition}
This proposition establishes that the counterfactual distribution of the treatment respondents is identified by evaluating the distribution of the untreated potential outcome of that subpopulation at baseline at the transformation $T_0(y)$ identified from Lemma \ref{lemma:CiC}. The ATT-R is then identified from the counterfactual distribution.

Next, we provide the identification result for the ATE-R, which requires the strict monotonicity of both treated and untreated potential outcomes in $U_t$. 
\begin{proposition}[Identification of the ATE-R]\label{prop:ATE-R}
Suppose that Assumptions \ref{ass:Unobs} and \ref{ass:Mon} hold. Suppose further that $\mathbb{U}_{0,1}=\mathbb{U}_{1,1}$. Then,

\begin{align}\text{ATE-R}=P(G=1|R=1)\text{ATT-R}+P(G=0|R=1)\text{ATU-R},
\end{align}
where
\begin{align}
    &\text{ATT-{R}}=E[Y_{1}|G=1,R=1]-E[F_{Y_{1}|G=0,R=1}^{-1}(F_{Y_{0}|G=0,R=1}(Y_{0}))|G=1,R=1]\nonumber\\
    &\text{ATU-R}=E[F_{Y_{1}|G=1,R=1}^{-1}(F_{Y_{0}|G=1,R=1}(Y_{0}))|G=0,R=1]-E[Y_{1}|G=0,R=1].\label{eq:CR_ATE}
\end{align}

\end{proposition}
The proof of the above proposition follows from Lemma \ref{lemma:CiC}. Since the ATE-R is a probability-weighted average of the ATT-R and ATU-R, the identification result in Proposition \ref{prop:ATE-R} builds on the identification of the ATT-R in Proposition \ref{prop:TR}. It then establishes the identification of the ATU-R, which requires identifying the treated potential outcome distribution for the control respondents. That distribution is obtained by evaluating the baseline distribution of control respondents at the transformation $T_1(y)$ identified from Lemma \ref{lemma:CiC}.

\subsubsection{Identification Results for the Study Population} 
In this section, we present identification results for the study population.  Since the random assignment of treatment simplifies the identification of the ATE, we provide identification results with and without that assumption.  Under random assignment, the identification of the ATE only requires identifying the treated (untreated) potential outcome distributions for treatment (control) attritors. Thus, researchers analyzing data from a randomized controlled trial can implement the correction indicated by Proposition \ref{prop:ATE_RA}. In contrast, researchers using other research designs should implement the correction indicated in Proposition \ref{prop:ATE} as the identification of the ATE relies on separately identifying the counterfactuals for the ATT and ATU for respondents and attritors.

We first examine the attrition correction without assuming random assignment. The law of iterated expectations allows us to write $E[Y_1(d)]$ as follows:
{\footnotesize{\begin{eqnarray*}
E[Y_1(d)] &=& P(G=1,R=1) E[Y_1(d)|G=1,R=1] + P(G=0,R=1) E[Y_1(d)|G=0,R=1]\\
&& + P(G=1,R=0) E[Y_1(d)|G=1,R=0] + P(G=0,R=0) E[Y_1(d)|G=0,R=0].
\end{eqnarray*}}}
\vspace{-0.8cm}

\noindent For $d=0$, the only terms that are observable on the right-hand side are the probabilities as well as the expected potential outcome without the treatment for the control respondents, $E[Y_1(0)|G=0,R=1]$.  Therefore, in order to identify $E[Y_1(0)]$, it remains to identify the distributions of the untreated potential outcome for all remaining subpopulations, $F_{Y_1(0)|G=1,R=0}$, $F_{Y_1(0)|G=1,R=1}$, and $F_{Y_1(0)|G=0,R=0}$.  Similarly, for $d=1$, the only terms that are observable on the right-hand side are the probabilities as well as the expected potential outcome with the treatment for the treatment respondents, $E[Y_1(1)|G=1,R=1]$.  As a result, in order to identify $E[Y_1(1)]$, it remains to identify  the distribution of the treated potential outcome for all remaining subpopulations, $F_{Y_1(1)|G=1,R=0}$, $F_{Y_1(1)|G=0,R=1}$, and $F_{Y_1(1)|G=0,R=0}$. 

The next proposition provides sufficient conditions such that we can apply Lemma \ref{lemma:CiC}.\ref{T0} to identify $F_{Y_1(0)|G=1,R=0}$, $F_{Y_1(0)|G=1,R=1}$, and $F_{Y_1(0)|G=0,R=0}$ as well as Lemma \ref{lemma:CiC}.\ref{T1} to identify $F_{Y_1(1)|G=1,R=0}$, $F_{Y_1(1)|G=0,R=1}$, and $F_{Y_1(1)|G=0,R=0}$. The identification of the ATE follows.

\begin{proposition}[Identification of the ATE without Random Assignment]\label{prop:ATE}  Suppose  Assumptions \ref{ass:Unobs} and \ref{ass:Mon} hold. Suppose further that $\mathbb{U}_{g,r}=\mathbb{U}$ $\forall (g,r)\in\{0,1\}^2$.\\
Then, \\
{\footnotesize{\begin{align}ATE=&P(R=1,G=1)\text{ATT-R}+P(R=0,G=1)\text{ATT-A}+P(R=1,G=0)\text{ATU-R}+P(R=0,G=0)\text{ATU-A},\nonumber\end{align}}}
where\\
{\footnotesize{$\text{ATT-R}=E[Y_1|G=1,R=1]-E[F_{Y_{1}|G=0,R=1}^{-1}(F_{Y_{0}|G=0,R=1}(Y_{0}))|G=1,R=1]$,\\
$\text{ATT-A}=E[F_{Y_{1}|G=1,R=1}^{-1}(F_{Y_{0}|G=1,R=1}(Y_{0}))|G=1,R=0]-E[F_{Y_{1}|G=0,R=1}^{-1}(F_{Y_{0}|G=0,R=1}(Y_{0}))|G=1,R=0]$,\\
$\text{ATU-R}=E[F_{Y_{1}|G=1,R=1}^{-1}(F_{Y_{0}|G=1,R=1}(Y_{0}))|G=0,R=1]-E[Y_{1}|G=0,R=1]$,\\
$\text{ATU-A}=E[F_{Y_{1}|G=1,R=1}^{-1}(F_{Y_{0}|G=1,R=1}(Y_{0}))|G=0,R=0]-E[F_{Y_{1}|G=0,R=1}^{-1}(F_{Y_{0}|G=0,R=1}(Y_{0}))|G=0,R=0]$.}}
\end{proposition}

Proposition \ref{prop:ATE} has two main practical implications.  First, it demonstrates that the CiC approach can identify the ATE in settings without (simple) random assignment, such as quasi-experimental difference-in-difference designs. We specifically have to obtain the average treatment effect for each treatment-response subgroup. For the treatment (control) respondents, we obtain the ATT-R (ATU-R) by applying the CiC approach to identify their average outcome without (with) the treatment. Furthermore, since we do not observe either potential outcome for the attritors, we have to apply the CiC approach to identify the average potential outcome with and without the treatment. The ATE is then obtained as a probability-weighted average of the group-specific average treatment effects.\\

Next, we examine the identification of the ATE under random assignment. Under this assumption, we have $ATE=E[Y_1(1)|G=1]-E[Y_1(0)|G=0]$. Using the law of iterated expectations, we have 
{\footnotesize{\begin{eqnarray*}
E[Y_1(0)|G=0] &=& P(R=1|G=0)E[Y_{1}|G=0,R=1]+P(R=0|G=0)E[Y_{1}(0)|G=0,R=0],\\
E[Y_1(1)|G=1]&=&P(R=1|G=1)E[Y_{1}|G=1,R=1]+P(R=0|G=1)E[Y_{1}(1)|G=1,R=0].
\end{eqnarray*}}}
The only unobservable objects on the right-hand side of the above equations are the average outcomes of the control and treatment attritors, $E[Y_{1}(0)|G=0,R=0]$ and $E[Y_{1}(1)|G=1,R=0]$.  The following proposition provides sufficient conditions such that we can apply Lemma \ref{lemma:CiC}.\ref{T0} and \ref{lemma:CiC}.\ref{T1}  to identify $F_{Y_{1}(0)|G=0,R=0}$ and $F_{Y_{1}(1)|G=1,R=0}$, respectively, and thereby their expectations.  

\begin{proposition}[Identification of the ATE under Random Assignment]\label{prop:ATE_RA}  Suppose Assumptions \ref{ass:Unobs}, \ref{ass:Mon} and \ref{ass:RA} hold. Suppose further that $\mathbb{U}_{0,1}=\mathbb{U}_{1,0}$, $\mathbb{U}_{1,0}\subseteq\mathbb{U}_{1,1}$,  $\mathbb{U}_{0,0}\subseteq\mathbb{U}_{0,1}$. Then,
{\footnotesize{\begin{align}ATE=&P(R=1|G=1)E[Y_{1}|G=1,R=1]+P(R=0|G=1)E[Y_{1}(1)|G=1,R=0]\nonumber\\
&-\left(P(R=1|G=0)E[Y_{1}|G=0,R=1)+P(R=0|G=0)E[Y_{1}(0)|G=0,R=0]\right)\nonumber
\end{align}
where\\ $E[Y_{1}(1)|G=1,R=0]=E[F_{Y_{1}|G=1,R=1}^{-1}(F_{Y_{0}|G=1,R=1}(Y_{0}))|G=1,R=0]$,\\
$E[Y_{1}(0)|G=0,R=0]=E[F_{Y_{1}|G=0,R=1}^{-1}(F_{Y_{0}|G=0,R=1}(Y_{0}))|G=0,R=0]$.}}
\end{proposition}
This proposition recovers the ATE in the case of random assignment by identifying the outcome distribution at follow-up of control attritors and respondents  using $T_0(y)$ and  $T_1(y)$ from Lemma \ref{lemma:CiC}, respectively. As a result, random assignment simplifies the identification of the ATE. The following remark demonstrates that it also provides a testable implication of the CiC assumptions.
\begin{remark}[Testable Implication of CiC Assumptions under Random Assignment] By establishing identification of the average treatment effects for the treatment-response subgroups in the absence of random assignment, Proposition \ref{prop:ATE} allows us to test the CiC assumptions in the presence of random assignment since random assignment (Assumption \ref{ass:RA}) implies that $ATT=ATU$. 

Under the CiC assumptions, we can separately identify the ATT and ATU by Proposition \ref{prop:ATE} as follows,
\begin{align}ATT_{CiC}&=ATT\text{-}R*P(R=1|G=1)+ATT\text{-}A*P(R=0|G=1)\\
ATU_{CiC}&=ATU\text{-}R*P(R=1|G=0)+ATU\text{-}A*P(R=0|G=0)\end{align}
As a result, under random assignment, we can test the CiC assumptions (Assumptions \ref{ass:Unobs}-\ref{ass:Mon}) as well as the support condition in Proposition \ref{prop:ATE} by testing the equality, $$ATT_{CiC}=ATU_{CiC}.$$ One of the advantages of this testable implication is that the magnitude of the violation of the CiC condition is measured in terms of differences between the average treatment effects of two subpopulations, which is easy to interpret from a practitioner's perspective. 
It is important to note, however, that this equality is not the sharp testable restriction of random assignment under the CiC conditions. The sharp testable restriction is indeed $F_{Y_1(d)|G=1}=F_{Y_1(d)|G=0}$ for all $d=0,1$, and $F_{Y_0(0)|G=1}=F_{Y_0(0)|G=0}$, where $F_{Y_1(d)|G=g}=P(R=1\vert G=g)F_{Y_1(d)|R=1,G=g}+P(R=0\vert G=g)F_{Y_1(d)|R=0,G=g}$. Developing a formal test of this sharp testable restriction is outside the scope of the present paper.\label{rem:test}
\end{remark}
\begin{remark}[Stratified Randomization]
For stratified randomized experiments, where $$(Y_0(0),Y_1(0),Y_1(1),R(0),R(1))\perp G|S,$$ and $S$ is the strata variable, empirical researchers have two potential avenues to identify the ATE.\footnote{In this case, researchers can still use Proposition \ref{prop:ATE-R} to identify the ATE-R since this proposition does not rely on random assignment.} They can implement the CiC correction exploiting random assignment within each stratum as in Proposition \ref{prop:ATE_RA}, assuming the CiC conditions hold conditional on strata (see Remark \ref{rem:covariates}). Alternatively, they can use the identification approach in Proposition \ref{prop:ATE}, which does not require random assignment, and therefore they can avoid conditioning on strata. Empirical researchers should consider the plausibility of the different identifying assumption for each approach as well as the relative number of strata to sample size in choosing between these different options. \label{rem:stratified}
\end{remark}

\begin{remark}[Covariates]
The identification results incorporating covariates follow in a straightforward manner assuming the suitable strict monotonicity and conditional time-invariance assumptions hold. Formally, suppose that the time-invariance condition holds conditional on $X$, $U_0|G,R,X\overset{d}{=}U_1|G,R,X$, and the strict monotonicity condition holds after incorporating a subset of $X$ in the potential outcome model. Then, we would have to implement the corrections conditional on $X$ following Propositions \ref{prop:ATE-R} and \ref{prop:ATE}.\footnote{If we, in addition, assume conditional random assignment as in stratified random assignments, $(Y_0(0),Y_1(0),Y_1(1),R(0),R(1))\perp G|S$, then we could implement the correction for the ATE under conditional random assignment following Proposition \ref{prop:ATE_RA}, where we condition on $(X,S)$.}

If the covariates are discrete, then our CiC corrections can be implemented on strata defined by the covariates. For instance, in stratified randomization designs, researchers can implement the corrections within each strata assuming the CiC conditions hold within treatment-response-stratum subgroups. For continuous covariates, there are several existing approaches in the literature. \citet{AI2006} present two approaches, a simple parametric approach that requires separability as well as a nonparametric approach that is difficult to implement in practice and suffers from the curse of dimensionality. \citet{MS2015} propose a flexible semiparametric approach relying on conditional quantile regressions.\footnote{For settings where quantile treatment effects are of interest, \citet{SasakiWang2023} propose a new CiC estimator for extreme quantiles including those conditioning on covariates.}

\label{rem:covariates}\end{remark}

\section{Comparison with Existing Attrition Corrections \label{subsec:CIC_comps}}
In this section, we describe the most widely-used corrections in practice, which are IPW and \citet{L2007} bounds, and compare their assumptions to the CiC assumptions.\footnote{See \citet{MM2017} for a review of approaches to correcting for attrition bias in the field experiment literature.} While the CiC correction exploits restrictions on the outcome model, the existing approaches exploit restrictions on how response depends on treatment.

\subsection{IPW Corrections}\label{sec:ipw}

Unlike the CiC approach, the IPW corrections rely on the assumption of selection on observables (i.e, unconfoundedness). In particular, to identify the average treatment effect on the respondents, it is required that treatment assignment is independent of potential outcomes and potential response, once we condition on baseline covariates $X_0$ ($G \perp (Y_1(0),Y_1(1),R(0),R(1)) \vert X_0$).\footnote{For the special case where $X_0=Y_0$, $G\perp (Y_1(0),Y_1(1),R(0),R(1))|Y_0$.} A second assumption required for the identification of the ATE-R is that potential response does not depend on treatment status, $R(0)=R(1)$.\footnote{While unconfoundedness by itself is not testable, combining it with $R(0)=R(1)$ implies  $G\perp (Y_1(0),Y_1(1),R)\vert X_0$, which further implies the testable restriction, $R\perp G\vert X_0$. This is therefore a testable restriction of the IPW identifying assumptions of the ATE-R. We emphasize, however, that the additional restriction required for the identification of the ATE relying on IPW, specifically $(R(0),R(1))\perp (Y_1(0),Y_1(1))|X_0$, is not testable.} This condition rules out individuals who would only respond if assigned to the treatment or control group, the so-called treatment-only or control-only responders, and implies that the respondents ($R=1$) solely consist of always-responders. In addition to unconfoundedness, the identification of the ATE requires that potential response is independent of potential outcome conditional on $X_0$, $(R(0),R(1))\perp (Y_1(0),Y_1(1))|X_0$.  See Section \ref{app:other_corrections} in the online appendix for the definitions of the ATE-R and ATE using these corrections.

There are two main differences between the IPW and CiC assumptions for the identification of the ATE-R. First, while the CiC approach exploits monotonicity of the outcome model, IPW restricts the response model and rules out the possibility of differential attrition rates across treatment and control groups, prevalent in the empirical literature.\footnote{In a detailed review of published field experiments \citet{GHO2022} find that 37\% (11\%) of field experiments have differential attrition rates higher than 2\% (5\%). This proportion is likely a lower bound on the proportion among all field experiments, given the publication bias towards field experiments with smaller differential attrition rates. It is important to note, however, that if treatment-only and control-only responders exist in the population, then differential attrition rates estimate the difference in proportion between these two responder subpopulations and, therefore, should not be used as an indication of whether response depends on treatment status or not. } In addition, the IPW approach requires treatment status to be conditionally randomly assigned, while the CiC correction allows for selection into response and treatment status under the condition that the distribution of unobservables $U_t$ within each treatment-response subgroup does not change between baseline and follow-up. Furthermore, in order to identify the ATE, IPW requires a conditional independence assumption between potential responses and outcomes, which states that missingness (i.e. attrition) is random conditional on $X_0$.

In sum, while the IPW and CiC approaches are non-nested in general, the main advantages of the CiC approach are twofold.  First, it allows response to depend on treatment, a likely concern in practice.  In addition, it does not require missingness to be conditionally at random to identify the ATE. Thus, there are a number of settings in which CiC can be applied where it would not be appropriate to apply IPW.  In settings where the assumptions of both approaches are suitable, however, we note that CiC requires the availability of non-degenerate baseline outcome data while IPW can be implemented when only baseline covariates are available. Furthermore, we note that IPW delivers point-identification of the ATE-R and ATE regardless of the outcome distribution.  In contrast, the CiC approach provides point-identification for continuous outcomes, and provides bounds on the ATT-R, ATE-R and ATE for discrete outcomes.

\subsubsection{CiC, IPW, and Response Models\label{subsec:resp_apps}}

When comparing CiC and IPW, it is helpful to consider specific response models such as those described in Section \ref{sec:response}. Example \ref{ex:prop4} provides a simple response function in the context of random assignment that can illuminate the implications of CiC conditions: $R=1\{V\geq c_0+c_1G\}$.  Here, we consider settings in which this type of response function may apply. Let $V$ be the opportunity cost of time, since that is likely an important reason that participants are reluctant to respond in many settings. We return to our example in which the treatment is a microcredit program, the goal of which is to increase profits.  As in the discussion of the high-level assumptions above, $U_t$ is an unobservable that is likely to affect a participant's work in the business and thus profits.  Since we are now also imposing structure on the response function, we note that the (normalized) health shock ($U_t$) would also affect the opportunity cost of time and thus the likelihood of responding to a survey.  That is, $V$ has a dependence relationship with $U_t$. In this case, the CiC assumptions allow the realization of the health shock and the opportunity cost of time to be related, as long as the relationship is stable over time.  In contrast, the IPW assumptions for the ATE-R require that the opportunity cost of time at follow-up only depends on the health status at baseline ($U_{0}$) rather than the health status at the time of the actual follow-up survey ($U_1$).

Alternatively, now let us consider two cases of the same example in which profits are instead largely determined by an unobservable, such as ability, which is constant over time ($U_0=U_1$).  First, consider the case in which $c_1>0$.  This may describe the relevant response function if the microcredit program allows beneficiaries to expand their businesses, and thus treated individuals are busier and less likely to respond. Of course, in these cases, it would not be appropriate to apply IPW, since the IPW assumptions do not allow response to vary with the treatment status. This case would meet the conditions for CiC, however, as the conditional time-invariance assumption holds trivially if $U_0=U_1$.

Next, let us consider a case in which migration determines response, and $V$ is a one-time shock at endline caused by a conflict.  The conflict shock could differentially affect participants based on ability, if, for example, higher ability individuals are better able to adapt and are less likely to leave.  In that case, $U$ and $V$ would be related. But, if the response function is the same for the treatment and control ($c_1=0$) and ability is constant over time, both the CiC and IPW assumptions (for the ATE-R and ATE) would hold.\footnote{We emphasize that this conclusion relies on random assignment. Furthermore, note that if we condition on $Y_0$ in IPW, then the conditional missing-at-random assumption required for the IPW correction for the ATE trivially holds when $U_0=U_1$ under Assumption \ref{ass:Mon}. To see this, note that under this assumption conditioning on $Y_0$ is equivalent to conditioning on $U$, which is assumed to be time-invariant in this example. As a result, $(R(0),R(1))\perp (Y_1(0),Y_1(1))|Y_0$ holds trivially, since the potential outcomes are fixed in this case once we condition on $U$.}

Now suppose that the unobservables that affect the outcome are the same unobservables that affect response ($U=V$). Corollary \ref{cor:CIC_selection} establishes a sufficient condition under which the time-invariance assumption and this restriction holds.  For example, consider a case in which the treatment is a matching grant program for charitable donations and the outcome is donations, and thus the preference for reciprocity is the (time-invariant) unobservable that determines the outcome as well as response.  In addition, treatment status may affect whether an individual will affect response to a survey ($c_1>0$). Since the preference for reciprocity is constant over time, the CiC assumptions would hold, whereas the IPW assumptions would not because response depends on treatment status.

To further examine the sufficient condition established by this corollary, we consider the response equation given in Example \ref{ex:cor1}, $R=1\{U_0+U_1\leq c_0+c_1G\}$.  This function describes a situation in which response depends on the flow of an unobservable that accumulates in each period. Let $U$, for example, be confidence.  Once a participant's confidence reaches a certain threshold, they migrate and are not available to respond to the survey. In addition, returning to our microcredit example, increasing confidence leads to greater business success and higher profits. It would make sense for confidence in the baseline period to matter, if the migrant needs to begin reaching out through their networks or otherwise planning for their migration in the baseline period. If the accumulated confidence across both the baseline and follow-up periods matter, then the CiC condition will hold and the IPW assumption will not. By contrast, if only the confidence accumulated in the baseline period matters, then the CiC assumptions would not hold in general, whereas the IPW assumptions for the ATE-R would only hold if $c_1=0$.  Finally, if only the additional confidence accumulated in the follow-up period matters, then both the CiC and IPW assumptions would fail in general.  

These examples highlight a range of possible examples of response functions, and how the CiC and IPW assumptions apply.\footnote{Although each of the examples above focuses on a single possible unobservable determining response for illustrative purposes, our approach allows for more complex response models as indicated in Example \ref{ex:prop4}.}  We focus on comparing the CiC and IPW assumptions since they both can recover the ATE-R and ATE. It is important to emphasize that to identify the ATE, IPW requires an additional assumption, specifically that attrition is random conditional on the covariates as we describe in Section \ref{sec:ipw}.

\subsection{Bounding Approaches}\label{subsec:bounds}

\citet{L2007}, aware of the likely possibility that some individuals are induced to respond due to their treatment status, exploits an assumption of monotonicity of response in treatment status, $R(0)\leq R(1)$, while maintaining unconfoundedness, $G \perp (Y_1(0),Y_1(1), R(0), R(1))|X_0$. This monotonicity assumption states that treatment assignment only affects response in one direction, and implies that the control respondents consist solely of always-responders $((R(0),R(1))=(1,1))$. Thus, as a result, it allows for the partial identification of the average treatment effect for the always-responders, a subset of the respondent subpopulation.\footnote{See Section \ref{app:other_corrections} in the online appendix for the equations with the bounds for continuous and binary outcomes.}  

The Lee bounds and the CiC corrections we propose in this paper differ in terms of the identifiable objects and the required conditions. First, \citet{L2007} bounds the average treatment effect for always-responders, which is a subpopulation of the respondents, whereas the CiC corrections we propose can identify the average treatment effects for the respondents as well as the entire study population. Second, \citet{L2007} requires random assignment of the treatment unconditionally or conditional on some covariates, whereas the CiC corrections can be extended to settings where treatment is not randomly assigned assuming that the distribution of the unobservable determinant of the outcome is stable over time for each treatment-response subgroup. Third, our approach requires monotonicity of the outcome in a scalar unobservable, while the \citeauthor{L2007}~(\citeyear{L2007}) approach requires instead monotonicity of the response in the treatment. This assumption is unlikely to hold in settings where nonresponse is determined by multiple factors, such as reciprocity or cost of time, since they are likely to lead to treatment-only responders and control-only responders, respectively.  One key advantage of the Lee bounds, however, is that they do not require baseline data. 

While Lee bounds are worst-case scenario bounds in the spirit of \citet{HM1995}, more recently, \citet{BCGB2015} exploit an insight that additional data on reluctance to respond to surveys combined with the same assumption of monotonicity of response can provide bounds that are tighter than \citeauthor{L2007}'s~(\citeyear{L2007}). To do so, the authors assume that the maximum number of attempts to reach an individual is randomly assigned and excluded from the outcome. In addition to monotonicity of response in treatment, they further require response to be monotone in the survey effort.

\section{Empirical Illustration}\label{sec:empirical}

We apply our proposed CiC corrections and other common corrections to two outcomes from a large-scale randomized evaluation of the impact of \textit{Progresa}, a conditional cash transfer program in Mexico. The \textit{Progresa} evaluation, which was implemented in 1997, randomized 506 villages into a treatment group and a control group.  These villages were designed to be representative of a larger group of 6,396 eligible villages in Mexico. Thus, both the average treatment effect for the respondent subpopulation (ATE-R) and for the study population (ATE) are likely to be of interest in this setting. In the 320 treatment villages, families received a conditional cash transfer if they were below the given threshold on a poverty index and engaged in specific education and health-seeking behaviors. In the control villages, no households  were offered a transfer.  There is a vast literature that has studied a range of outcomes from the \textit{Progresa} evaluation, with some of the most studied outcomes focusing on education and health \citep{Skoufias2001,Schultz2004,AMS2012, PT2017}.

The goal of this application is to demonstrate the implementation of the CiC correction on a continuous outcome with baseline data, as well as the comparison of the CiC and IPW corrections for that outcome.\footnote{We only include baseline outcome in both the CiC and IPW corrections to simplify the direct comparison of approaches since the CiC correction requires baseline outcome data.  Both corrections allow covariates, however. The use of covariates in the CiC correction is discussed in Remark \ref{rem:covariates}.}  Both CiC and IPW provide point estimates for the ATE-R and the ATE for continuous outcomes, and IPW is the most widely-used correction in the literature.\footnote{Another widely used attrition correction are the bounds proposed by \citet{L2007}. Since this approach only focuses on the average treatment effect on the always-responders, however, it is not directly comparable to our method that focuses on the ATE-R and the ATE.} In Section \ref{app:empirical_bounds} of the online appendix, we also apply both the Manski and Lee bounding approaches to this continuous outcome, and implement all four corrections for a related binary variable. 

The continuous outcome we examine is the value of a productive asset, specifically farm animals. Close to 90\% of the households in the population targeted by \textit{Progresa} engage in agricultural activities and make investments in productive assets for their farms.\footnote{This outcome is first proposed in \citet{GMR2012}.  Our findings are not directly relevant to that paper since we only focus on the third follow-up, while \citet{GMR2012} focus on the outcome pooled across all three follow-ups.  We also restrict our sample to those who appear in the baseline survey.}  At baseline, the average value of farm animals in these households was 1,819 Mexican pesos (denoted \$), which is equivalent to approximately 102 USD today.

The potential for attrition bias in treatment effect estimates from the original \textit{Progresa} follow-ups has often been discussed in the literature  \citep{Bobonis2011, PRT2007, BT1999}. Thus, attrition in this setting is significant and warrants the implementation of corrections. We focus here on the final follow-up that takes place 18 months after the program began, and that has an attrition rate of 12.2\%.\footnote{This attrition rate is close to the average attrition rate for field experiments where the unit of observation is the individual or household \citep{MM2017}. We focus on this final follow-up since examining a single follow-up allows us to more clearly outline reasons for attrition, and the final follow-up is often seen as definitive.  Furthermore, since assets are not likely to adjust quickly, it makes sense to focus on impacts in the final follow-up. That said, for researchers who would like to generate pooled estimates of the CiC corrected estimates, they can simply average across the corrected estimates for the individual follow-ups.}  For the purposes of this analysis, the attrition rate is conditional on appearing in the baseline survey, which included a total of 12,299 households. Thus, the outcome is observed for more than 10,000 households in this follow-up survey.

\subsection{Results of Application}\label{sec:appresults}

We first examine the CiC-corrected estimates for the ATE-R and ATE in relation to the na\"{i}ve (uncorrected) estimate of the treatment effect for productive assets,  $\widehat{\Delta}_R$, which is simply the difference in the mean outcome between treatment and control respondents at follow-up.  The na\"{i}ve estimate of the treatment effect on the value of production animals is \$351, which is significant at the 1\% level and is relative to a control mean of \$1,096 (see Panel A of Table \ref{tab:empirical}).  The CiC-corrected estimate for the ATE-R is \$284 while the CiC-corrected estimate for the ATE is \$289. The similarity of these two estimates suggests that, if the CiC assumptions hold, there is relatively little treatment heterogeneity. A key consideration, however, is whether these coefficients differ from the na\"{i}ve estimate of the treatment effect (Panel B of Table \ref{tab:empirical}).  On the one hand, the CiC-corrected ATE-R is not significantly different from the na\"{i}ve treatment effect estimate, since the difference in the two estimates is \$67 with a standard error of \$125.   On the other hand, for the ATE, the difference is \$62 with a standard error of \$26, which is significant at the 5\% level. A difference in the power of these tests may explain this result.

Meanwhile, the IPW approach does not suggest that a correction is required for either the ATE-R or the ATE. The IPW-corrected estimates, which are \$349 for the ATE-R and \$342 for the ATE, are nearly identical in magnitude to the na\"{i}ve estimate and are not close to being significantly different from it even at the 10\% level.  Thus, the CiC-corrected estimates for the ATE imply that the treatment effect is a 26\% increase relative to the control while the IPW-corrected estimates imply that the treatment effect is a 31\% increase.  This difference in two corrected estimates is modest, but statistically significant at the 5\% level and potentially meaningful in considering the returns to programs such as \textit{Progresa}.

\subsection{Identification Tests and their Implications} \label{sec:idtests}

Since the different corrections we implement rely on different identifying assumptions, it is crucial to discuss which are more plausible in this setting. In order to shed some light on this question, we first consider whether attrition is likely to be causing a violation of internal validity that would necessitate a correction. Thus, we implement the tests proposed in \citet{GHO2022} to assess the impact of attrition on the internal validity of the estimated treatment effects in the \emph{Progresa} study (see Panel C of Table \ref{tab:empirical}). These tests rely on a panel framework, and are based on testable restrictions of the relevant identifying assumptions in the presence of attrition that ensure internal validity for the respondents (IVal-R), which identifies the ATE-R, and the study population (IVal-P), which identifies the ATE. The testable restriction for  IVal-R consists of the equality of the mean baseline outcome across treatment and control groups conditional on response status, that is, $E[Y_{0}|G=0,R=r]=E[Y_{0}|G=1,R=r]$ for $r=0,1$. In contrast, the testable restriction for IVal-P consists of the equality of the mean baseline outcome across the four treatment-response subgroups, $E[Y_{0}|G=g,R=r]= E[Y_{0}]$ for $(g,r)\in\{0,1\}^2$.

In applying these tests to the productive asset outcome, we find that the test of internal validity for the respondents (IVal-R) is not rejected, whereas the test of interval validity for the study population (IVal-P) is rejected. Indeed, we find substantially higher mean baseline value of production animals for respondents relative to attritors. Thus, consistent with the findings of the CiC correction, the results of these tests for attrition bias indicate that a correction is required for the ATE but not the ATE-R.\footnote{We emphasize that while these tests are helpful in interpreting the corrections, they should not be used as pre-tests to decide whether to implement a correction or not due to the resulting pre-test bias issue.} A caveat, however, is that it is possible that the IVal-R test is unable to detect a violation of internal validity for respondents.

Next, we consider whether the CiC assumptions hold.  The CiC approach under attrition relies on two high-level assumptions for the four treatment-response subgroups: a monotonic relationship between the outcome and its unobservable determinant as well as a conditional time-invariant distribution for the unobservable in question.  Specifically, we apply the test of the implication of CiC assumptions under random assignment, $ATT_{CiC}=ATU_{CiC}$, (see Remark \ref{rem:test}). As shown in Panel D of Table \ref{tab:empirical}, the CiC estimates of the ATT and ATU are almost exactly equal with a difference of $0.1$ and this difference is not statistically significant.  Thus, since we do not find evidence that the CiC assumptions are violated here, the results of this test are consistent with interpreting the CiC corrections as estimates of causal impacts. 

By contrast, the IPW conditions do not allow response to depend on treatment status, which implies equal attrition rates. In this setting, the differential rate of 2.6 percentage points, but is marginally not significant with a p-value 0.117.\footnote{In Section \ref{app:ipw_assumpt} of the online appendix, we test the restriction of the two IPW assumptions required for the identification of the ATE-R. There, we marginally reject the equality of attrition rates across treatment and control groups, but do not reject the joint null that response is independent of treatment conditional on $Y_0$.}  Furthermore, equal attrition rates are not sufficient to ensure that response does not depend on treatment when monotonicity of response is violated.
  For the ATE, the IPW further requires the assumption that attrition is random conditional on variable(s) that are included in the correction, which is not testable.

\subsection{Plausibility of the Identifying Assumptions}\label{sec:appassumptions}

To complement the results of these tests of identifying assumptions, we also heuristically consider the plausibility of the CiC or the IPW assumptions in this setting by considering likely reasons for attrition. The IPW assumptions do not restrict the outcome model, but do restrict response.  Meanwhile the CiC conditions restrict the outcome model, but they are compatible with a wide range of response models.  If researchers have a specific response model in mind, they can apply that understanding in considering whether the conditional time-invariance assumption holds.\footnote{For more general comparisons of the CiC and IPW approaches under various response models, see Section \ref{subsec:resp_apps} and Section \ref{sec:response}.}

Other studies that propose corrections generally focus on one of two main reasons for attrition: participants may simply be reluctant to respond to interviews or migration may hamper the enumerators' ability to find and interview participants \citep{BCGB2015, MM2017}.  There are several factors that influence each of these reasons for attrition, however, and thus neither reason maps directly into a particular set of unobservables that  determine response. For example, migration can be triggered primarily by a high tolerance for risk driving a willingness to search for better economic opportunities or by covariate shocks such as conflict and droughts. Likewise, the main determinant of the reluctance to respond to surveys in any given setting could be one of a range of different types of factors: (lack of) reciprocity, the sensitivity of the questions, and the opportunity cost of time.

It is not common practice for researchers to report analysis on reasons for attrition in field experiments, and thus it is not surprising that specific data on the reasons for attrition in the \emph{Progresa} evaluation is not publicly available.\footnote{In conducting a review of attrition in 96 published field experiments, \citet{GHO2022} did not find that such studies discuss data on reasons for attrition in general. That is not surprising since collecting data on why it is not possible to find a specific respondent, when they cannot be found, may not be possible by construction.  Of course, in some cases it is possible to collect such data from a respondent's associates.  There may also be occasional circumstances in which general reasons for attrition may be understood without additional data collection, such as when civil unrest or a natural disaster drives displacement, even if such a reason cannot be linked to specific respondents. } Although stated reasons for attrition are not typically available, authors commonly examine drivers of attrition by implementing a determinants of attrition test that examines how respondents and attritors differ in terms of baseline covariates, which can help one infer what are the likely underlying unobservables determining response. In \citet{GHO2022}, we find that authors conduct a determinants of attrition test for 29\% of experiments where there is attrition. Thus, we implement such a test for this application using available baseline data (see Table \ref{tab:det_att} in the online appendix).

While examining covariates of attrition cannot reveal a specific underlying response function, it can suggest potential patterns of attrition that are consistent with the results of the application and the tests described in Section \ref{sec:idtests}.  For example, the findings from the determinants of attrition test are consistent with the idea that reluctance in the form of opportunity cost of time is a key reason for survey non-response. In particular, response is significantly correlated with the likelihood of having an adult at home or being a farm household, which is unsurprising given that households that engage in agricultural production are more likely to work at home and thus have a lower opportunity cost of time during the day to respond to a survey.\footnote{We define a farm household as one that used agricultural land or owned production animals at baseline, which differs from the outcome of value of production animals.  Of course, alternative drivers of attrition are consistent with the findings from the determinants of attrition test as is discussed in Section \ref{app:dat} of the online appendix.}  Response here then is plausibly related to an unobservable determinant of the outcome in this setting, $U_t$, which could represent the skills to succeed in accumulating agricultural assets, for example.  This would be consistent with the CiC assumptions if the joint distribution between agricultural skills and the opportunity cost of time is stable across time. Furthermore, since being a farm household likely depends on several unobservables that affect both response and potential outcomes at follow-up, we expect attrition to be nonrandom, even after conditioning on the baseline covariates. If that is the case, IPW assumptions for the ATE would not be satisfied.

Of course, relying on determinants of attrition tests have limitations, however, in fully capturing potential reasons for attrition.  In particular, baseline data is less likely to be informative about  stochastic factors, such as health shocks.  It is plausible that such a factor is an unobservable determinant of agricultural assets, and also explains a differential reluctance to respond across treatment and control groups in this setting, even if the test of differential attrition rates here is marginally insignificant. As discussed in Section \ref{subsec:resp_apps}, such a case can be accommodated by our model, but cannot be accommodated by the IPW assumptions.  As we also discuss in that section, the IPW assumptions and the CiC assumptions would both hold in this setting because it is a randomized experiment, if: (i) $U$ is some constant factor across time such as ability, and (ii) response is not affected by treatment status. In general, however, the most realistic models of response are likely to be those in which more than one factor influences response.  As discussed in Example \ref{ex:prop4}, such models can be accommodated by the CiC assumptions.

\section{Conclusion \label{concl}}
In this paper, we propose an attrition correction method for the average treatment effects on the respondents as well as the study population that relies on the CiC framework. We achieve identification through two main assumptions: continuity and strict monotonicity of each potential outcome in a scalar unobservable, and time invariance of the distribution of the unobservable conditional on treatment and response status. We then show that these assumptions are likely to hold in a range of typical settings, and can accommodate a variety of different response functions.  

We further compare these assumptions with other widely-used approaches.  We focus in particular on the comparison with the IPW correction, since it provides point-identification for the average treatment effect for respondents (ATE-R) and the average treatment effect from the study population (ATE).  The IPW correction relies on the assumptions of response independent of treatment status and unconfoundness for the ATE-R as well as conditionally random attrition for the ATE.  In contrast to the CiC approach, Lee bounds rely on monotonicity of response, and are not designed to provide bounds for the ATE.  We illustrate the performance and plausibility of these corrections for an application to an outcome from the randomized evaluation of the \textit{Progresa} conditional cash transfer program in Mexico. 

The CiC approach provides point-estimates for continuous outcomes and bounds for discrete outcomes.  Given researchers commonly consider both continuous and discrete or binary versions of the same outcome, this study highlights that there is potential value in focusing on continuous versions of outcomes when correcting for attrition.     
The CiC corrections proposed here do not require random assignment, but do require that baseline outcome is available.  Thus, they can be applied to quasi-experimental difference-in-difference designs. There is an ongoing debate, however, about the value of collecting baseline data for randomized controlled trials.  Of course, there are settings where the baseline outcome is degenerate by design, however when that is not the case, this paper points to the value of collecting baseline outcome data.

\newpage

\begin{table}[htbp]
\centering
\caption{Value of Production Animals (Mexican pesos)} 
\label{tab:empirical}
{\small{
\begin{tabular}{lccccccc}\\
\multicolumn{8}{c}{Panel A. Observed Difference in Means and Attrition Corrections}\\
\toprule
Estimator&\multicolumn{1}{c}{\parbox{2.3em}{Control Mean}} &$\widehat{\Delta}_R$& \multicolumn{3}{c}{{CiC}} & \multicolumn{2}{c}{{IPW}}  \\
\cmidrule(lr){4-6}\cmidrule(lr){7-8}
&&&ATT-R & ATE-R& ATE & ATE-R &ATE  \\
&(1)&(2)&(3)&(4)&(5)&(6)&(7)\\
 \midrule
Estimate & 1,096&  351.0*** & 285.6*** & 283.9*** & 288.9*** & 348.6*** & 342.2*** \\
S.E. & 86.0&   118.9 & 92.4 &94.6 &  107.1 & 113.3  &110.3\\
 
\bottomrule
\\
\end{tabular}\\

\begin{tabular}{lccccccc}
\multicolumn{8}{c}{Panel B. Differences between Estimates} \\
\toprule
Estimator &  \multicolumn{1}{c}{\parbox{3.2em}{\centering (2)-(4)}} &  \multicolumn{1}{c}{\parbox{3.2em}{\centering (2)-(5)}}& \multicolumn{1}{c}{\parbox{3.2em}{\centering (4)-(5)}} &\multicolumn{1}{c}{\parbox{3.2em}{\centering (4)-(6)}}&\multicolumn{1}{c}{\parbox{3.2em}{\centering (5)-(7)}} &\multicolumn{1}{c}{\parbox{3.2em}{\centering (2)-(6)}} &\multicolumn{1}{c}{\parbox{3.2em}{\centering (2)-(7)}}\\
  \midrule
  \multicolumn{1}{p{4em}}{Difference}&  67.1 & 62.0** & -5.1 & -64.7 & -53.2** &2.5 & 8.8 \\
  S.E.&  124.7 & 26.0 & 113.8 &116.6 &22.4 & 9.9 & 12.4 \\
  \bottomrule
\\    \end{tabular}\\

 \begin{tabular}{ccccccccc}
    \multicolumn{9}{c}{Panel C. Attrition Rates, Baseline Outcome, and Attrition Tests}\\
    \toprule
 \multicolumn{3}{c}{Attrition Rates} & \multicolumn{4}{c}{Mean Baseline Outcome by Group} & \multicolumn{2}{c}{Attrition Tests} \\
 \cmidrule(lr){1-3}
 \cmidrule(lr){4-7}\cmidrule(lr){8-9}
  Overall & T &T-C & TR    & CR    & TA    & CA    & IVal-R & IVal-P \\
    \midrule
12.2&13.2&2.6&1,905.7& 1,850.5 & 1,294.8 & 1,455.6 & 0.713 & 0.000 \\
    \bottomrule \\
    \end{tabular}\\
    
\begin{threeparttable}
\begin{tabular}{p{2cm} p{2cm} p{2cm} p{2cm}}
\multicolumn{4}{c}{Panel D. Testing CiC Assumptions under Random Assignment} \\
\toprule
Estimator &  ATT & ATU & ATT-ATU \\
  \midrule
Estimate &272.6***&272.5***&  0.1   \\
S.E.&88.2& 97.9&30.0   \\
  \bottomrule
  \end{tabular}
    \begin{tablenotes}[normal, flushleft]
			\item \small{ \emph{Notes}: This table reports the results of the CiC and IPW attrition corrections for the outcome of value of production animals in the third follow-up of the \emph{Progresa} evaluation. The IPW correction only includes the baseline outcome as a covariate. The number of households at baseline and follow-up are 12,299 and 10,799, respectively. Standard errors are bootstrapped. The attrition tests of internal validity for the respondent sub-population (IVal-R) and the study population (IVal-P) correspond to those proposed in \citet{GHO2022}. P-values are reported for these tests. \sym{***} $p<0.01$; \sym{**} $p<0.05$; \sym{*} $p<0.1$.}
		\end{tablenotes}
	\end{threeparttable}
}}
\end{table}

\newpage
\bibliographystyle{aer}
\bibliography{bibtex,did}

\newpage
\appendix

\section{Proof of the Main Results}\label{app:proofs}
\begin{proof} (Lemma \ref{lemma:CiC})

\noindent 1. For $(g,r)\in\{0,1\}^2$, Assumptions \ref{ass:Unobs} and \ref{ass:Mon}.\ref{MonUPO} imply the following for $y$ in the support of $Y_1(0)|G=g,R=r$
\begin{align}
    F_{Y_1(0)|G=g,R=r}(y)&=P(\mu_1(0,U_1)\leq y|G=g,R=r)=P(U_1\leq \mu_1^{-1}(0;y)|G=g,R=r)\nonumber\\
   & =P(U_0\leq \mu_1^{-1}(0;y)|G=g,R=r)= P(\mu_0(0,U_0)\leq \mu_0(0,\mu_1^{-1}(0;y))|G=g,R=r)\nonumber\\
    &=F_{Y_{0}(0)|G=g,R=r}(\mu_0(0,\mu_1^{-1}(0;y)))\equiv F_{Y_{0}(0)|G=g,R=r}(T_0(y))\quad for\quad y\in\mathbb{Y}_{g,r}^{0,1}\label{eq:F_Y01_Y00}
\end{align}
where the second equality holds by Assumption \ref{ass:Mon}.\ref{MonUPO}, the third equality follows from Assumption \ref{ass:Unobs}.\ref{TI} and the fourth equality follows by applying the transformation $\mu_0(0,.)$ which is valid by Assumption \ref{ass:Mon}.\ref{MonUPO}. The result in (i) follows by the definition of $T_0(y)=\mu_0(0,\mu_1^{-1}(0;y))$.

The identification of $T_0(y)$ on $y\in\mathbb{Y}_{0,1}^{0,1}$ follows from \eqref{eq:F_Y01_Y00}, the fact that we observe $F_{Y_t(0)|G=0,R=1}$ for $t=0,1$, and the strict monotonicity of $F_{Y_{0}(0)|G=0,R=1}(\cdot)$ by Assumptions \ref{ass:Unobs}.\ref{continuity} and \ref{ass:Mon}.\ref{MonUPO}, which imply
\begin{align}T_0(y)&=F_{Y_0|G=0,R=1}^{-1}(F_{Y_1|G=0,R=1}(y))\quad for\quad y\in\mathbb{Y}_{0,1}^{0,1}.
\end{align}
This completes the proof of (i) and (ii).
 
\noindent 2. By similar arguments, Assumptions \ref{ass:Unobs}, \ref{ass:Mon}.\ref{MonUPO} and \ref{ass:Mon}.\ref{MonTPO} imply the following for $(g,r)\in\{0,1\}^2$, 
\begin{align}
    F_{Y_1(1)|G=g,R=r}(y)&=P(\mu_1(1,U_1)\leq y|G=g,R=r)=P(U_1\leq \mu_1^{-1}(1;y)|G=g,R=r)\nonumber\\
   & =P(U_0\leq \mu_1^{-1}(1;y)|G=g,R=r)= P(\mu_0(0,U_0)\leq \mu_0(0,\mu_1^{-1}(1;y))|G=g,R=r)\nonumber\\
    &=P(Y_{0}(0)\leq \mu_0(0,\mu_1^{-1}(1;y))|G=g,R=r)\nonumber\\
    &    \equiv F_{Y_{0}(0)|G=g,R=r}(T_1(y))\quad for\quad y\in\mathbb{Y}_{g,r}^{1,1}\label{eq:F_Y11_Y00}
\end{align}

The identification of $T_1(y)$ on $y\in\mathbb{Y}_{1,1}^{1,1}$ follows from \eqref{eq:F_Y11_Y00}, the fact we observe $F_{Y_0(0)|G=1,R=1}$ and $F_{Y_1(1)|G=1,R=1}$, and the strict monotonicity of $F_{Y_{0}(0)|G=1,R=1}(\cdot)$ by Assumptions \ref{ass:Unobs}.\ref{continuity} and \ref{ass:Mon}.\ref{MonUPO}, which imply
\begin{align}T_1(y)&=F_{Y_0|G=1,R=1}^{-1}(F_{Y_1|G=1,R=1}(y))\quad for\quad y\in\mathbb{Y}_{1,1}^{1,1}.
\end{align}
This completes the proof of (i) and (ii).
\QEDB

\end{proof}

\begin{proof} (Proposition \ref{prop:CIC_selection})\\
The proof relies on the following identities. Note that since $P(R=r|G)\neq 0$ for $r=0,1$, then for $t=0,1$ and $u \in \mathbb R$
\begin{align}P(U_{t}\leq u|G,R=r)&=\frac{P(U_{t}\leq u,R=r|G)}{P(R=r|G)}
\end{align}
Hence, time homogeneity of $U_{t}$ given $G$ and $R$ holds iff $P(U_{t}\leq u,R=r|G)$ is time homogeneous. We have
\begin{align}
    P(U_{t}\leq u,R=r|G=g)&=P(U_{t}\leq u, \varphi(g,V)=r|G=g),\nonumber \\
    &= P(U_{t}\leq u, V \in \varphi^{-1}(g,r)|G=g), \label{eq:homogeneous}
\end{align}
where $\varphi^{-1}(g,r)=\{v:\varphi(g,v)=r\}$. 

Therefore, 
\begin{align*}
    P(U_{0}\leq u,R=r|G=g)&= P(U_{0}\leq u, V \in \varphi^{-1}(g,r)|G=g),\\
    &= P(U_{1}\leq u, V \in \varphi^{-1}(g,r)|G=g),\\
    &= P(U_{1}\leq u,\varphi(g,V)=r|G=g),\\
    &= P(U_{1}\leq u,\varphi(G,V)=r|G=g),\\
    &=  P(U_{1}\leq u,R=r|G=g)
\end{align*}
where the second equality follows by the time-invariance of $(U_t,V)|G$, the third holds from the definition of $\varphi^{-1}$, and the last holds because $R=\varphi(G,V)$. 

Now, suppose $U_{0}|G,R\overset{d}{=}U_{1}|G,R$. Define $V:=R$. Then $U_{0}|G,V\overset{d}{=}U_{1}|G,V$, which implies $(U_{0},V)|G \overset{d}{=} (U_{1},V)|G$  since $V$ is time-invariant. Define $\varphi(g,v):= v$ for all $(g,v)$. Clearly, the function $\varphi$ is measurable as it is continuous, $R=\varphi(G,V)$ by construction, and $V$ is a random vector of one dimension. 

This completes the proof.\QEDB
\end{proof}

\begin{proof} (Corollary \ref{cor:CIC_selection})\\

Under the assumptions of Corollary \ref{cor:CIC_selection}, we have $R=\psi(G,U_0,U_1)=\psi(G,U_1,U_0)$. As in the proof of Proposition \ref{prop:CIC_selection}, we need to show that $P(U_{0}\leq u,R=r|G=g)=P(U_{1}\leq u,R=r|G=g)$. 
Define $\psi^{-1}(g,r)=\{(u_0,u_1): \psi(g,u_0,u_1)=r\}$.
We have
\begin{align*}
    P(U_{0}\leq u,R=r|G=g)&= P(U_{0}\leq u,\psi(g,U_0,U_1)=r|G=g),\\
    &= P(U_{0}\leq u,(U_0,U_1) \in \psi^{-1}(g,r)|G=g),\\
    &= P((U_0,U_1) \in ((-\infty,u] \times \mathbb R) \cap \psi^{-1}(g,r)|G=g) ,\\
    &=P((U_1,U_0) \in ((-\infty,u] \times \mathbb R) \cap \psi^{-1}(g,r)|G=g) ,\\
    &= P(U_{1}\leq u,(U_1,U_0) \in \psi^{-1}(g,r)|G=g),\\ 
    &=  P(U_{1}\leq u,\psi(g,U_1,U_0)=r|G=g),\\
    &= P(U_{1}\leq u,R=r|G=g),
\end{align*}
where the fourth equality holds under exchangeability, and the last holds under symmetry and the fact that $U_0$ and $U_1$ follow the same distribution. 
This completes the proof

\QEDB
\end{proof}

\begin{proof} (Proposition \ref{prop:TR}) \\
 Since all conditions required for Lemma \ref{lemma:CiC}.\ref{T0} are imposed, $T_0(y)$ is identified for $y\in\mathbb{Y}_{0,1}^{0,1}$ (Lemma \ref{lemma:CiC}.\ref{T0}(ii)).   The imposed support condition, $\mathbb{U}_{1,1}\subseteq\mathbb{U}_{0,1}$, together with the strict monotonicity of the untreated potential outcome (Assumption \ref{ass:Mon}.\ref{MonUPO}) implies that $\mathbb{Y}_{1,1}^{0,1}\subseteq \mathbb{Y}_{0,1}^{0,1}$.  As a result, $T_0(y)$ is identified for $y\in\mathbb{Y}_{1,1}^{0,1}$.  By Lemma \ref{lemma:CiC}.\ref{T0}(i)-(ii), the following equality in \eqref{eq:TR_counterfactual} follows,
\begin{align}F_{Y_1(0)|G=1,R=1}(y)&=F_{Y_0|G=1,R=1}(F_{Y_0|G=0,R=1}^{-1}(F_{Y_1|G=0,R=1}(y)))\quad for \quad y\in\mathbb{Y}_{1,1}^{0,1}.\end{align} The identification of the ATT-R in \eqref{eq:TR_ATE} is immediate from \eqref{eq:TR_counterfactual}.
\QEDB
\end{proof}

\begin{proof} (Proposition \ref{prop:ATE-R})\\
The identification of the ATT-R follows by Proposition \ref{prop:TR}. It follows to provide the identification of the ATU-R. Since all conditions required for Lemma \ref{lemma:CiC}.\ref{T1} hold, $T_1(y)$ is identified for $y\in\mathbb{Y}_{1,1}^{1,1}$.  The support condition imposed here, $\mathbb{U}_{0,1}\subseteq\mathbb{U}_{1,1}$, together with the strict monotonicity of the treated potential outcome (Assumption \ref{ass:Mon}.\ref{MonTPO}) implies that $\mathbb{Y}_{0,1}^{1,1}\subseteq \mathbb{Y}_{1,1}^{1,1}$.  As a result, $T_1(y)$ is identified for $y\in\mathbb{Y}_{0,1}^{1,1}$. By Lemma \ref{lemma:CiC}.\ref{T1}(i)-(ii), it follows that 
\begin{align}F_{Y_1(1)|G=0,R=1}(y)&=F_{Y_0|G=0,R=1}(F_{Y_0|G=1,R=1}^{-1}(F_{Y_1|G=1,R=1}(y)) \quad for\quad y\in\mathbb{Y}_{0,1}^{1,1}.\end{align}
The identification of the ATU-R in \eqref{eq:CR_ATE} is immediate from the identification of the distribution of $Y_1(1)|G=0,R=1$.

\QEDB
\end{proof}

\begin{proof} (Proposition \ref{prop:ATE}) 
We have for all $d=0,1$,
{\footnotesize{\begin{eqnarray*}
E[Y_1(d)] &=& P(G=1,R=1) E[Y_1(d)|G=1,R=1] + P(G=0,R=1) E[Y_1(d)|G=0,R=1]\\
&& + P(G=1,R=0) E[Y_1(d)|G=1,R=0] + P(G=0,R=0) E[Y_1(d)|G=0,R=0].
\end{eqnarray*}}}
Then 
{\footnotesize{\begin{eqnarray*}
E[Y_1(1)]-E[Y_1(0)] &=& P(G=1,R=1) E[Y_1(1)-Y_1(0)|G=1,R=1] + P(G=0,R=1) E[Y_1(1)-Y_1(0)|G=0,R=1]\\
&& + P(G=1,R=0) E[Y_1(1)-Y_1(0)|G=1,R=0] + P(G=0,R=0) E[Y_1(1)-Y_1(0)|G=0,R=0].
\end{eqnarray*}}}
Under the maintained assumptions, the conditions for Lemma \ref{lemma:CiC} hold. Therefore, the distributions $F_{Y_1(1)|G=1,R=0}$, $F_{Y_1(1)|G=0,R=1}$, $F_{Y_1(1)|G=0,R=0}$, $F_{Y_1(0)|G=1,R=0}$, $F_{Y_1(0)|G=1,R=1}$, and $F_{Y_1(0)|G=0,R=0}$ are identified. The distributions, $F_{Y_1(1)|G=1,R=1}$ and $F_{Y_1(0)|G=0,R=1}$, as well as the probability weights, $P(G=g,R=r)$ for all $g=0,1$ and $r=0,1$, are identified from the data. Therefore, all objects in the definition of the ATE given in the proposition are identified.  It follows that the ATE is identified.
\QEDB
\end{proof}

\begin{proof} (Proposition \ref{prop:ATE_RA})
Note that by random assignment
\begin{align}
    &E[Y_{1}(1)-Y_{1}(0)] =E[Y_{1}(1)|G=1]-E[Y_{1}(0)|G=0],\nonumber\\
    =&P(R=1|G=1)E[Y_{1}(1)|G=1,R=1]+P(R=0|G=1)E[Y_{1}(1)|G=1,R=0]\nonumber\\
    &-P(R=1|G=0)E[Y_{1}(0)|G=0,R=1]-P(R=0|G=0)E[Y_{1}(0)|G=0,R=0].\nonumber
\end{align}
All of the quantities on the RHS of the last equality are observed except $E[Y_{1}(1)|G=1,R=0]$ and $E[Y_{1}(0)|G=0,R=0]$.  Under the maintained assumptions, the conditions for Lemma \ref{lemma:CiC} hold. Therefore, we can identify the distribution of $Y_{1}(1)$ for the treatment attritors as well as the distribution of $Y_{1}(0)$ for the control attritors.  Specifically, we have
\begin{align*}
&F_{Y_{1}(1)|G=1,R=0}(y)=F_{Y_{0}|G=1,R=0}(F_{Y_{0}|G=1,R=1}^{-1}(F_{Y_{1}|G=1,R=1}(y))),\\
&E[Y_{1}(1)|G=1,R=0]=E[F_{Y_{1}|G=1,R=1}^{-1}(F_{Y_{0}|G=1,R=1}(Y_{0}))|G=1,R=0],
\end{align*}
and
\begin{align*}
&F_{Y_{1}(0)|G=0,R=0}(y)=F_{Y_{0}|G=0,R=0}(F_{Y_{0}|G=0,R=1}^{-1}(F_{Y_{1}|G=0,R=1}(y))),\\
&E[Y_{1}(0)|G=0,R=0]=E[F_{Y_{1}|G=0,R=1}^{-1}(F_{Y_{0}|G=0,R=1}(Y_{0}))|G=0,R=0].\end{align*}
As a result, the ATE is identified.
\QEDB
\end{proof}

\newpage

\vspace{5cm}
\begin{center}\vspace{2cm}
\huge{Correcting Attrition Bias using Changes-in-Changes}\\
\medskip
\Large{Dalia Ghanem \hspace{0.3cm} Sarojini Hirshleifer \hspace{0.3cm} D\'esir\'e K\'edagni \hspace{0.3cm} Karen Ortiz-Becerra\\
\medskip

Online Appendix\\
\date{\today}}\end{center}
\vspace{1cm}

\setcounter{page}{1}
\setcounter{table}{0}
\renewcommand\theHtable{Appendix.\thetable} 
\renewcommand{\thetable}{SA\arabic{table}}
\setcounter{figure}{0}
\renewcommand{\thefigure}{SA\arabic{figure}}
\setcounter{section}{0}
\renewcommand{\thesection}{SA\arabic{section}}

\doublespacing

\numberwithin{equation}{section}
\vspace{-0.25cm}

 \titlecontents{section}[1.5em]
  {\normalfont}
   {\contentslabel{2em}} 
   {\hspace*{-2.3em}}
   {\titlerule*[1pc]{.}\contentspage}

 \titlecontents{subsection}[4.5em]
  {\normalfont}
    {\contentslabel{3em}}
    {\hspace*{-2.3em}}
    {\titlerule*[1pc]{.}\contentspage}
  
 \startcontents[sections]
 \printcontents[sections]{l}{1}{\setcounter{tocdepth}{2}}

\newpage

\section{Attrition Corrections Using Difference-in-Differences Approaches}\label{app:DiD}

Given the simplicity and wide use of difference-in-differences, a natural question that arises is what types of identifying conditions would be required for us to identify the same objects of interests using difference-in-differences. 
Assuming that parallel trends holds conditional on response, \[E[Y_1(0)-Y_0(0)|G,R]=E[Y_1(0)-Y_0(0)|R]\tag{\text{PT Conditional on Response}}\] then we can identify the ATT-R using difference-in-differences by the following
\begin{align}E[Y_1(1)-Y_1(0)|G=1,R=1]&=E[Y_1-Y_0|G=1,R=1]-E[Y_1-Y_0|G=0,R=1]\end{align}
However, the parallel trends assumption conditional on response would not be sufficient to identify the remaining objects of interest, specifically ATU-R, ATT-A, ATU-A, unless we assume constant treatment effects.\footnote{It is also worth noting that even a strong parallel trends condition $E[Y_1(0)-Y_0(0)|G,R]=E[Y_1(0)-Y_0(0)]$ would not be sufficient to identify those objects. It would only be sufficient to identify the average untreated potential outcome for the attritors $E[Y_1(0)|G=g,R=0]=E[Y_0|G=g,R=0]-E[Y_1(0)-Y_0(0)|G=0,R=1]$ for $g=0,1$. Since it does not restrict the treated potential outcome, we cannot rely on it to identify the ATU-R, ATT-A and ATU-A.}

In the following, we provide an alternative model that allows for heterogeneity in the treatment effect in a restrictive way that would allow us to identify the ATU-R, ATT-A and ATU-A in addition to the ATT-R:
\begin{align}Y_t&=\alpha(1+b D_{t})+\lambda_t+\varepsilon_{t}(1+b D_{t}),\label{eq:did2}\end{align}
where $D_t$ denotes treatment status in period $t$ ($D_t \equiv G\cdot t$), $\alpha$ is a time-invariant unobservable, $\lambda_t$ is assumed to be nonstochastic, $b$ is a constant, and $\varepsilon_t$ represents time-varying unobservables. If we assume $E[\varepsilon_0|G,R]=E[\varepsilon_1|G,R]$, then the parallel trends assumption holds for all groups $(G,R)$: 
\begin{align}&E[Y_1(0)-Y_0(0)|G,R]=\lambda_1-\lambda_0.\end{align}
Since we can only identify $\lambda_1-\lambda_0$, we can normalize $\lambda_0=0$ without loss of generality. Under this normalization and $E[\varepsilon_0|G,R]=E[\varepsilon_1|G,R]$, the following equality holds:
\begin{align}
E[Y_0|G,R]&=E[\alpha+\varepsilon_0|G,R].\label{eq:alpha+epsilon_id}\end{align}
From Equation \eqref{eq:did2}, we have 
\begin{align}Y_{t}(1)=\alpha+b\alpha
+\lambda_t+(1+b)\varepsilon_{t},\ \text{ and }\ Y_t(0)=\alpha + \lambda_t + \varepsilon_t 
\end{align}
such that $Y_1(1)-Y_1(0)=b(\alpha+\varepsilon_1)$, where $b$ is a constant. Here the ATT-R$=b E[\alpha+\varepsilon_1|G=1,R=1]$. Now note that we can identify $E[Y_0|G=g,R=r]=E[\alpha+\varepsilon_0|G=g,R=r]=E[\alpha+\varepsilon_1|G=g,R=r]$ from \eqref{eq:alpha+epsilon_id} and $E[\varepsilon_0|G,R]=E[\varepsilon_1|G,R]$. Assuming $E[Y_0|G=g,R=r]\neq 0$, then
\begin{align*}
\text{ATU-R}&=\text{ATT-R}\frac{E[Y_0|G=0,R=1]}{E[Y_0|G=1,R=1]},\\
\text{ATT-A}&=\text{ATT-R}\frac{E[Y_0|G=1,R=0]}{E[Y_0|G=1,R=1]},\\
\text{ATU-A}&=\text{ATT-R}\frac{E[Y_0|G=0,R=0]}{E[Y_0|G=1,R=1]}.
\end{align*}
As a result, under the restriction on the untreated potential outcome, we can further identify the ATU-R, ATT-A, ATU-A as the ratio of the average treatment effects for the different subgroups are proportional to their average outcome at baseline. This allows for only specific types of heterogeneity in the treatment effect. Furthermore, noting that $U_t=\alpha+\varepsilon_t$, the above model satisfies the monotonicity restrictions in the scalar unobservable $U_t$ required by the changes-in-changes approach. In practice, however, researchers would not want to take a particular stance on the functional form and therefore the advantage of the CiC approach is that it can handle any monotonic function. Furthermore, the transformation in the CiC approach is time-varying in an arbitrary way beyond location shifts, not only for the model of the untreated potential outcome but also the treated potential outcome, and thereby the treatment effects. 

\section{Bounding Average Treatment Effects for Discrete Outcomes}\label{app:discrete_extension}
While we focus our identification results on continuous outcomes, following \citet{AI2006} we can provide bounds for discrete outcomes. Define $F^{(-1)}_{Y}(q)=\sup\left\{y \in \mathbb Y \cup \{-\infty\}: F_Y(y) \leq q\right\}$.

\begin{proposition}\label{prop:discrete_outcomes} 
Suppose that Assumption \ref{ass:Unobs} in the paper holds. Suppose further that $\mu_t(d,u)$ is nondecreasing in $u$ for $d=0,1$ and $\mathbb{U}_{g,r}=\mathbb{U}$ $\forall (g,r)\in\{0,1\}^2$, then for $(g,r)\in\{0,1\}^2$ {\footnotesize{\begin{eqnarray*}(i)& F_{Y_0\vert G=g,R=r}\left(F_{Y_0\vert G=0,R=1}^{(-1)}(F_{Y_1\vert G=0,R=1}(y))\right)\leq  F_{Y_1(0)\vert G=g,R=r}(y)\leq   F_{Y_0\vert G=g,R=r}\left(F_{Y_0\vert G=0,R=1}^{-1}(F_{Y_1\vert G=0,R=1}(y))\right).\\
(ii)& F_{Y_0\vert G=g,R=r}\left(F_{Y_0\vert G=1,R=1}^{(-1)}(F_{Y_1\vert G=1,R=1}(y))\right)\leq  F_{Y_1(1)\vert G=g,R=r}(y)\leq   F_{Y_0\vert G=g,R=r}\left(F_{Y_0\vert G=1,R=1}^{-1}(F_{Y_1\vert G=1,R=1}(y))\right).\\
\end{eqnarray*}}}
\end{proposition}
Proposition \ref{prop:discrete_outcomes} provides bounds on the potential outcome distributions with and without the treatment at follow-up, respectively. The bounds on the ATT-R, the ATE-R and the ATE can be obtained relying on these bounds, as we show in Section \ref{app:empirical_bounds_binary} of this online appendix.
\begin{proof} The proof of (i) and (ii) will rely on the following inequality for discrete outcome distribution $F_Y(\cdot)$ from AI2006
\begin{eqnarray}&F_{Y}(F_Y^{(-1)}(q))\leq q\leq F_Y(F_Y^{-1}(q))\label{eq:AI2006inequality}\end{eqnarray}

\noindent (i) We first proceed to show two useful equalities,
\begin{eqnarray}
    F_{Y_t\vert G=0,R=1}(y) &=& \mathbb P(\mu_t(0,U_t)\leq y|G=0,R=1),\nonumber\\
    &=& F_{U_t\vert G=0,R=1}(\sup\{u:\mu_t(0,u)=y\}),\nonumber\end{eqnarray}
    where the last equality follows from $\mu_t(0,u)$ being non-decreasing in $u$.
    \begin{eqnarray}
    F_{Y_t(0)\vert G=g,R=r}(y) &=& \mathbb P(Y_t(0) \leq y \vert G=g,R=r),\nonumber\\
    &=&  \mathbb P(\mu_t(0,U_t) \leq y \vert G=g,R=r),\nonumber\\
    &=&  \mathbb P(U_t \leq \sup\{u:\mu_t(0,u)=y\} \vert G=g,R=r),\nonumber\\
    &=& \mathbb P(F_{U_t\vert G=0,R=1}(U_t) \leq F_{U_t\vert G=0,R=1}(\sup\{u:\mu_t(0,u)=y\}) \vert G=g,R=r),\nonumber\\
    &=& \mathbb P(\tilde{U}^{01}_t \leq F_{Y_t\vert G=0,R=1}(y) \vert G=g,R=r) \text{ from the above result},\nonumber\\
    &=& F_{\tilde{U}_t^{01}\vert G=g,R=r}(F_{Y_t\vert G=0,R=1}(y)),\label{eq:Y(0)_equality2}
\end{eqnarray}
where $\tilde{U}^{01}_t=F_{U_t\vert G=0,R=1}(U_t)$.

Now we proceed to show the lower bound in (i)
\begin{eqnarray*}
   && F_{Y_0\vert G=g,R=r}\left(F_{Y_0\vert G=0,R=1}^{(-1)}(F_{Y_1\vert G=0,R=1}(y))\right)\\
   && \qquad \qquad = F_{\tilde{U}_0^{01}\vert G=g,R=r}\left(F_{Y_0\vert G=0,R=1}\left(F_{Y_0\vert G=0,R=1}^{(-1)}(F_{Y_1\vert G=0,R=1}(y))\right)\right), \text{ as } Y_0(0)=Y_0,\\
    && \qquad \qquad = \mathbb P\left(\tilde{U}_0^{01} \leq \left(F_{Y_0\vert G=0,R=1}\left(F_{Y_0\vert G=0,R=1}^{(-1)}(F_{Y_1\vert G=0,R=1}(y))\right)\right)\vert G=g,R=r\right),\\
    && \qquad \qquad \leq \mathbb P\left(\tilde{U}_0^{01} \leq F_{Y_1\vert G=0,R=1}(y)\vert G=g,R=r\right),\\
    && \qquad \qquad = \mathbb P\left(\tilde{U}_1^{01} \leq F_{Y_1\vert G=0,R=1}(y)\vert G=g,R=r\right)\ \text{ under Assumption \ref{ass:Unobs}.\ref{TI}},\\
    && \qquad \qquad = F_{Y_1(0)\vert G=g,R=r}(y).
\end{eqnarray*}
where the first equality follows from \eqref{eq:Y(0)_equality2} and the inequality follows from the lower bound in \eqref{eq:AI2006inequality}.

Similarly, using the upper bound in \eqref{eq:AI2006inequality}, we provide the upper bound in (i).
\begin{eqnarray*}
   && F_{Y_0\vert G=g,R=r}\left(F_{Y_0\vert G=0,R=1}^{-1}(F_{Y_1\vert G=0,R=1}(y))\right)\\
   && \qquad \qquad = F_{\tilde{U}_0^{01}\vert G=g,R=r}\left(F_{Y_0\vert G=0,R=1}\left(F_{Y_0\vert G=0,R=1}^{-1}(F_{Y_1\vert G=0,R=1}(y))\right)\right),\\
    && \qquad \qquad = \mathbb P\left(\tilde{U}_0^{01} \leq \left(F_{Y_0\vert G=0,R=1}\left(F_{Y_0\vert G=0,R=1}^{-1}(F_{Y_1\vert G=0,R=1}(y))\right)\right)\vert G=g,R=r\right),\\
    && \qquad \qquad \geq \mathbb P\left(\tilde{U}_0^{01} \leq F_{Y_1\vert G=0,R=1}(y)\vert G=g,R=r\right),\\
    && \qquad \qquad = \mathbb P\left(\tilde{U}_1^{01} \leq F_{Y_1\vert G=0,R=1}(y)\vert G=g,R=r\right)\ \text{ under Assumption \ref{ass:Unobs}.\ref{TI}},\\
    && \qquad \qquad = F_{Y_1(0)\vert G=g,R=r}(y).
\end{eqnarray*}
\noindent (ii) The proof of (ii) proceeds in similar steps to (i).
\begin{eqnarray*}
    F_{Y_t\vert G=1,R=1}(y) &=& \mathbb P(\mu_t(1,U_t)\leq y|G=1,R=1),\\
    &=& F_{U_t\vert G=1,R=1}(\sup\{u:\mu_t(1,u)=y\})\ \text{ if } \mu_t(1,u)\ \text{ is nondecreasing in }u,\\
    F_{Y_t(1)\vert G=g,R=r}(y) &=& \mathbb P(Y_t(1) \leq y \vert G=g,R=r),\\
    &=&  \mathbb P(\mu_t(1,U_t) \leq y \vert G=g,R=r),\\
    &=&  \mathbb P(U_t \leq \sup\{u:\mu_t(1,u)=y\} \vert G=g,R=r),\\
    &=& \mathbb P(F_{U_t\vert G=1,R=1}(U_t) \leq F_{U_t\vert G=1,R=1}(\sup\{u:\mu_t(1,u)=y\}) \vert G=g,R=r),\\
    &=& \mathbb P(\tilde{U}^{11}_t \leq F_{Y_t\vert G=1,R=1}(y) \vert G=g,R=r) \text{ from the above result},\\
    &=& F_{\tilde{U}_t^{11}\vert G=g,R=r}(F_{Y_t\vert G=1,R=1}(y)),
\end{eqnarray*}
where $\tilde{U}^{11}_t=F_{U_t\vert G=1,R=1}(U_t)$.

\begin{eqnarray*}
   && F_{Y_0\vert G=g,R=r}\left(F_{Y_0\vert G=1,R=1}^{(-1)}(F_{Y_1\vert G=1,R=1}(y))\right)\\
   && = F_{\tilde{U}_0^{11}\vert G=g,R=r}\left(F_{Y_0\vert G=1,R=1}\left(F_{Y_0\vert G=1,R=1}^{(-1)}(F_{Y_1\vert G=1,R=1}(y))\right)\right), \text{ as } Y_0=Y_0(0),\\
    && = \mathbb P\left(\tilde{U}_0^{11} \leq \left(F_{Y_0\vert G=1,R=1}\left(F_{Y_0\vert G=1,R=1}^{(-1)}(F_{Y_1\vert G=1,R=1}(y))\right)\right)\vert G=g,R=r\right),\\
    && \leq \mathbb P\left(\tilde{U}_0^{11} \leq F_{Y_1\vert G=1,R=1}(y)\vert G=g,R=r\right),\\
    && = \mathbb P\left(\tilde{U}_1^{11} \leq F_{Y_1\vert G=1,R=1}(y)\vert G=g,R=r\right)\ \text{ under Assumption \ref{ass:Unobs}.\ref{TI}},\\
    && = F_{Y_1(1)\vert G=g,R=r}(y).
\end{eqnarray*}

Similarly,
\begin{eqnarray*}
   && F_{Y_0\vert G=g,R=r}\left(F_{Y_0\vert G=1,R=1}^{-1}(F_{Y_1\vert G=1,R=1}(y))\right)\\
   && \qquad \qquad = F_{\tilde{U}_0^{11}\vert G=g,R=r}\left(F_{Y_0\vert G=1,R=1}\left(F_{Y_0\vert G=1,R=1}^{-1}(F_{Y_1\vert G=1,R=1}(y))\right)\right),\\
    && \qquad \qquad = \mathbb P\left(\tilde{U}_0^{11} \leq \left(F_{Y_0\vert G=1,R=1}\left(F_{Y_0\vert G=1,R=1}^{-1}(F_{Y_1\vert G=1,R=1}(y))\right)\right)\vert G=g,R=r\right),\\
    && \qquad \qquad \geq \mathbb P\left(\tilde{U}_0^{11} \leq F_{Y_1\vert G=1,R=1}(y)\vert G=g,R=r\right),\\
    && \qquad \qquad = \mathbb P\left(\tilde{U}_1^{11} \leq F_{Y_1\vert G=1,R=1}(y)\vert G=g,R=r\right)\ \text{ under Assumption \ref{ass:Unobs}.\ref{TI}},\\
    && \qquad \qquad = F_{Y_1(1)\vert G=g,R=r}(y).
\end{eqnarray*}
\QEDB
\end{proof}

\section{Identification of Treatment Effects Using Alternative Approaches}
\label{app:other_corrections}

\subsection{IPW Corrections}
The IPW approach requires the following two assumptions to identify the average treatment effect on the respondents:
\begin{enumerate}
    \item $R(0)=R(1)$
    \item $G \perp (Y_1(0),Y_1(1),R(0),R(1)) \vert X_0$
\end{enumerate}

where $X_0$ is a vector of baseline variables that may contain the baseline outcome $Y_0$.\footnote{For the special case where $X_0=Y_0$, the unconfoundedness assumption is given by $G\perp (Y_1(0),Y_1(1),R(0),R(1))|Y_0$.}  The first assumption is that potential response does not depend on treatment status, while the second assumption states that selection into treatment is as if it were randomly assigned once we condition on response and covariates (i.e., unconfoundedness).

Under these assumptions, the ATE-R is identified as:
\begin{align}\text{ATE-R}&=E\left[\frac{Y_1 G}{P(G=1\vert R=1,X_0)}-\frac{Y_1 (1-G)}{1-P(G=1\vert R=1,X_0)}\vert R=1\right],\label{eq:ATE-R_IPW_X_0}\end{align}

If, in addition to unconfoundedness, we assume that potential response is independent of potential outcome conditional on $X_0$, $(R(0),R(1))\perp (Y_1(0),Y_1(1))|X_0$, then the IPW correction identifies the average treatment effect for the study population as:
\begin{align}\text{ATE}&=E\bigg[\frac{Y_1GR}{P(G=1|R=1,X_0)P(R=1|X_0)}-\frac{Y_1(1-G)R}{(1-P(G=1|R=1,X_0))P(R=1|X_0)}\bigg].\label{eq:ATE_IPW_X_0}\end{align}

\subsection{Lee Bounds}

\citet{L2007} exploits an assumption of monotonicity of response in treatment status, $R(0)\leq R(1)$, while maintaining conditional random assignment, $G \perp (Y_1(0),Y_1(1), R(0), R(1))|X_0$. Under the monotonicity of response, the control respondents consist solely of always-responders ($(R(0),R(1))=(1,1)$), whereas the treatment respondents consist of both always-responders and treatment-only responders. In what follows, we provide the formulas that identify the lower and upper Lee bounds for the subset of always-responders. For the sake of clarity of exposition, we do not condition on $X_0$.

By random assignment, the proportion of always-responders and treatment-only responders are the same across treatment and control groups, it follows that the proportion of treatment-only responders equals the difference in attrition rates between treatment and control groups,
\begin{align*}P((R(0),R(1))=(0,1))=P(R=1|G=1)-P(R=1|G=0).\end{align*}

Assuming that the outcome $Y$ is continuous, \citet{L2007} proposes worst-case scenario bounds on the potential outcome with the treatment for always-responders ($(R(0),R(1))=(1,1))$ as follows
\begin{eqnarray*}
E[Y_1(1)\vert (R(0),R(1))=(1,1)] &\in&\Bigg[ E\left[Y_1\vert R=1, G=1, Y_1 \leq F_{Y_1\vert R=1, G=1}^{-1}\left(\alpha\right)\right],\\
&& E\left[Y_1\vert R=1, G=1, Y_1 > F_{Y_1\vert R=1, G=1}^{-1}\left(1-\alpha\right)\right]\Bigg],
\end{eqnarray*}
where $\alpha=\frac{P(R=1\vert G=0)}{P(R=1\vert G=1)}$.
Since the control respondents solely consist of always-responders under the monotonicity of response, $E[Y_1\vert R=1, G=0] = E[Y_1(0)\vert R(0)=1]=E[Y_1(0)\vert (R(0),R(1))=(1,1)]$, where the first equality holds under random assignment, and the second holds under monotonicity. It follows that the sharp bounds derived by \citet{L2007} on the average treatment effect for the always-responders are given by
\begin{eqnarray}
&&E[Y_1(1)-Y_1(0)\vert (R(0),R(1))=(1,1)]\\ \nonumber
&\in&\Bigg[ E\left[Y_1\vert R=1, G=1, Y_1 \leq F_{Y_1\vert R=1, G=1}^{-1}\left(\alpha\right)\right]-E[Y_1|G=0,R=1],\\ \nonumber
&& \qquad \qquad E\left[Y_1\vert R=1, G=1, Y_1 > F_{Y_1\vert R=1, G=1}^{-1}\left(1-\alpha\right)\right]-E[Y_1|G=0,R=1]\Bigg].
\end{eqnarray}

When the outcome is binary, the worst-case bounds are obtained by imputing the
missing outcome at follow-up with all 1’s or all 0’s appropriately \citep{L2002}. In this case, the bounds on the average treatment effect for the always-responders are given by\begin{eqnarray}\label{eq:Lee_binary}
&&E[Y_1(1)\vert (R(0),R(1))=(1,1)]\\ \nonumber
&\in &\Bigg[\max\left\{0, \frac{\mathbb P(Y_1=1\vert R=1, G=1)+\alpha-1}{\alpha}\right\},\min\left\{1, \frac{\mathbb P(Y_1=1\vert R=1, G=1)}{\alpha}\right\}\Bigg].
\end{eqnarray}

\section{Supplementary Empirical Analysis \label{app:empirical_bounds}}

\subsection{Testable Restriction of IPW Assumptions for ATE-R}\label{app:ipw_assumpt}

In this section, we report the results of the testable restriction for the IPW assumptions to identify the ATE-R in the empirical application in Section \ref{sec:empirical}. As described in Section \ref{sec:ipw}, this correction requires unconfoundedness and  $R(0)=R(1)$, which together imply the testable restriction, $R\perp G|X_0$.  In the case in which $X_0=Y_0$, this restriction can be tested by estimating the model $R=\beta_0 +\beta_1 G + \beta_2 Y_0 + \beta_3 G\times Y_0 +V$, and testing the null hypothesis $H_0: \beta_1 = \beta_3 =0$, where $R$ refers to the outcome-specific response at follow-up, $G$ refers to treatment status, and $Y_0$ refers to the baseline outcome.

Table \ref{tab:ipw_assumpt} reports the results of this analysis for the continuous outcome of value of production animals, and shows that we do not reject the joint hypothesis that response is independent of treatment conditional on $Y_0$.

\begin{table}[htbp]
\centering
\caption{Testable Restriction of IPW Assumptions for ATE-R} 
\label{tab:ipw_assumpt}
{\small{
\begin{threeparttable}
\begin{tabular}{lcc}
\toprule
&\multicolumn{2}{c}{Response (=1)}\\
\cmidrule(lr){2-3}
 & (1)       & (2)     \\  
 \midrule
Treatment (=1) & -0.149*   &   -0.029*    \\
& (0.09)    &   (0.02)    \\
Value of production animals at baseline (\$)   &  0.000         &  0.000* \\
&    (0.00)        &  (0.00)     \\
Treatment * Value of production animals at baseline   & 0.000  & 0.00    \\
& (0.00)    &    (0.00)    \\
Constant & 1.227*** & 0.890*** \\
&(0.07) & (0.01) \\
\midrule
Testable restriction (p-val) & 0.228 & 0.217 \\
Observations         & 24,598 & 24,598 \\ 
\bottomrule
 \end{tabular}
    \begin{tablenotes}[normal, flushleft]
			\item \small{ \emph{Notes}: This Table reports the results of the testable restriction of the IPW  assumptions to identify the ATE-R for the outcome of value of production animals in the third follow-up of the \emph{Progresa} evaluation. Column (1) reports the results of the probit model, and column (2) reports the coefficients of the linear regression. Standard errors are clustered at the locality level. \sym{***} $p<0.01$; \sym{**} $p<0.05$; \sym{*} $p<0.1$.}
		\end{tablenotes}
	\end{threeparttable}
}}
\end{table}

\subsection{Determinants of Attrition}\label{app:dat}

In this section, we examine the predictors of response for the outcome of value of production animals from the empirical application in Section \ref{sec:appassumptions}.  Specifically, we report the results of a determinants of attrition test, an approach that is widely used in the literature, which relies on baseline covariates to predict response at follow-up. We focus on covariates in the available data that are likely to be proxies for the opportunity cost of time or predictors of migration.

First, we examine likely proxies for the opportunity cost of time that may explain whether someone in the household is available to answer the follow-up survey (Table \ref{tab:det_att}).\footnote{We examine the relationship between the outcome variable of interest (value of production animals), attrition and treatment using the appropriate, formal tests in Section \ref{sec:empirical}.  They are included in this analysis simply as controls rather than so that the coefficients can be interpreted.} In particular, households are less likely to respond if the head is female and more likely to respond if the household size is large. Being a farm household, which we define as using agricultural land or owning production animals,  is also a significant predictor of response at follow-up. This is expected given that households that engage in agricultural activities are more likely to work at home.  

When we examine likely predictors of migration, we find that neither wealth nor previous migration history are significant predictors of attrition. Household head education, however, is positively correlated with the availability of follow-up data.  This is not surprising if international migrants to countries that are relatively close geographically are those with lower levels of education.

One interpretation of this analysis is that the opportunity cost of time is a key reason for nonresponse in this setting relative to migration.  As discussed in Section \ref{sec:empirical} of the paper, however, the exercise of mapping results from a determinants of attrition test to drivers of attrition is likely to be suggestive rather than definitive.  In this setting, for example, it is possible that some of the proxies for the cost of time are also related to migration. In particular, farm households may be less likely to migrate in the short term due to the ownership of agricultural assets. 

\begin{table}[htbp]
\centering
\caption{Determinants of Attrition Analysis} 
\label{tab:det_att}
{\small{
\begin{threeparttable}
\begin{tabular}{lccc}
\toprule
&\multicolumn{3}{c}{Response (=1)}\\
\cmidrule(lr){2-4}
 & (1)       & (2)       & (3)        \\
 \midrule
Treatment (=1) & -0.003    & -0.005    & -0.005     \\
& (0.02)    & (0.02)    & (0.02)      \\
Value of production animals (\$)   &           & 0.000***  & 0.000***    \\
&           & (0.00)    & (0.00)      \\
Female household head (=1)   & -0.027**  & -0.027**  & -0.023*    \\
& (0.01)    & (0.01)    & (0.01)      \\
Number of children 0-16 yrs old & 0.013***  & 0.012***  & 0.010***   \\
& (0.00)    & (0.00)    & (0.00)      \\
Number of members $>$ 16 yrs old & 0.020***  & 0.017***  & 0.019***   \\
& (0.00)    & (0.00)    & (0.00)      \\
Farm household (=1)   &           & 0.052***  & 0.054***   \\
&           & (0.01)    & (0.01)      \\
Household head with high education level (=1) &           &           & 0.013**     \\
&           &           & (0.01)      \\
Wealth index   &           &           & -0.000    \\
&           &           & (0.00)      \\
Former member migrated within last 5 years (=1)   &           &           & 0.007       \\
    &           &           & (0.02)      \\
constant   & 0.787***  & 0.750***  & 0.775***   \\
& (0.01)    & (0.02)    & (0.04)     \\
\midrule
Municipality FE      & Yes       & Yes       & Yes         \\
Adj. R2              & 0.104     & 0.107     & 0.108       \\
Observations         & 12,278& 12,278 & 12,242 \\ 
\bottomrule
 \end{tabular}
    \begin{tablenotes}[normal, flushleft]
			\item \small{ \emph{Notes}: This Table reports the results of the determinants of attrition analysis for the outcome of value of production animals in the third follow-up of the \emph{Progresa} evaluation. All the covariates refer to characteristics of the household at baseline. A farm household is defined as one that uses agricultural land or owns production animals. High education level takes the value of one if the number of education years is above the median in the sample. The wealth index refers to the proxy means index used to classify poor vs. nonpoor households in the evaluation. Standard errors are clustered at the locality level. \sym{***} $p<0.01$; \sym{**} $p<0.05$; \sym{*} $p<0.1$.}
		\end{tablenotes}
	\end{threeparttable}
}}
\end{table}

\subsection{Manski and Lee Bounds for The Continuous Outcome \label{app:empirical_bounds1}}

To complement the empirical application in Section \ref{sec:empirical} of the paper, we also apply two bounding approaches to the continuous outcome from the evaluation of the \textit{Progresa} program we study, which is the value of production animals.  First, we apply the bounds proposed in \citet{Manski1989}.  This approach provides worst-case bounds for the average treatment effect for the subpopulation of respondents (ATE-R) and the average treatment effect for the study population (ATE) by replacing the missing data with the minimum or maximum value of production animals at follow-up. As shown in Table \ref{tab:empirical_bounds1}, the bounds for the ATE-R and the ATE are [\$-36034, \$23271] and [\$-38872, \$27665], respectively, and thus are inconclusive regarding the sign of the treatment effects in this setting. These wide bounds are not surprising given that the value of production animals among respondents ranges from \$0 to \$59,304. We also apply the Lee bounds to our data, an approach that is prevalent in empirical practice and provides bounds on the average treatment effect for the always-responders. According to this approach, the average treatment effect for the always responders is between \$318 and \$714. 

These bounding approaches are alternatives to the CiC and IPW corrections, which provide point estimates of the ATE and the ATE-R (see Section \ref{sec:empirical} of the paper).  In addition, in this application, both of those corrections provide estimates that are positive and statistically different from zero. An essential consideration for a researcher in choosing a correction, however, is whether the assumptions are appropriate for their setting.  The plausibility of the underlying assumptions for IPW and CiC corrections are discussed in Section \ref{sec:appassumptions} of the paper. Manski bounds require the assumption that the counterfactual outcomes have bounded support and that the attritors' support is included in the respondents'.  In contrast to the other approaches, the Lee approach bounds the average treatment effect for the always-responders, a subset of the respondent subpopulation.  It also requires the assumption that being assigned to treatment only affects response in one direction (i.e. monotonicity in response), ruling out a setting where, for example, receiving a cash transfer increases both reciprocity and the likelihood of migrating.

\begin{table}[htbp!]
\centering
\caption{Value of Production Animals (Mexican pesos)}
\label{tab:empirical_bounds1}
{\small{
\begin{threeparttable}
\begin{tabular}{lcccccccc}\\
\toprule
Estimator &$\widehat{\Delta}_R$& \multicolumn{1}{c}{{CiC}} & \multicolumn{1}{c}{{IPW}} & \multicolumn{2}{c}{\parbox{7.5em}{\centering Manski Bounds}} & \multicolumn{2}{c}{{Lee Bounds}} &\multicolumn{1}{c}{\parbox{5em}{\centering Tests of IVal}}\\
\cmidrule(lr){5-6}\cmidrule(lr){7-8}\cmidrule(lr){9-9}
& & & & LB & UB & LB & UB &$p$-val \\
\midrule
& (1) & (2) & (3) & (4) & (5) &(6) & (7) & (8) \\
\midrule
ATT-R  &  & 285.6***&&&&& & \\
ATE-R & 351.0***&283.9***&348.6***&-36,034&23,271&&&0.713  \\
ATE  & 351.0***&288.9***&342.2***&-38,872&27,665&&&0.000 \\
ATE - Always Responders &&&&&& 318.1 & 714.3& \\
\bottomrule \\
\end{tabular}
\begin{tablenotes}[normal, flushleft]
\item \small{ \emph{Notes}: This Table reports the results of the attrition corrections for the outcome of value of production animals in the third follow-up of the \emph{Progresa} evaluation. The number of households at baseline and follow-up are 12,299 and 10,799, respectively. Columns (4)-(7) report the unconditional version of the Manski and Lee bounds proposed in \citet{Manski1989} and \citet{L2007}, respectively. The Manski bounds were obtained by replacing the missing values with the minimum or maximum value of production animals among respondents at follow-up, which are \$0 and \$59,304.3, respectively. Column (8) reports the tests of internal validity proposed in \citet{GHO2022}. \sym{***} $p<0.01$; \sym{**} $p<0.05$; \sym{*} $p<0.1$.}
\end{tablenotes}
\end{threeparttable}
}}
\end{table}

\subsection{Partial Identification for An Alternative Binary Outcome}\label{app:empirical_bounds_binary}

The CiC approach provides bounds for discrete outcomes (Section \ref{app:discrete_extension}).  In this section, we illustrate the application of those bounds as well as alternative corrections. Specifically, we focus on the ownership of production animals in the third follow-up, which is the binary version of the continuous outcome (namely, the value of production animals) analyzed in Section \ref{sec:empirical} of the paper.  We focus on this particular binary variable since researchers often have a choice in focusing on a binary or continuous version of the same outcome.  This example highlights the potential gains of focusing on the continuous version of a variable when correcting for attrition bias is a priority.   

The attrition rates and patterns for the binary outcome analyzed in this section are identical to those discussed for the continuous outcome in Section \ref{sec:empirical} of the paper.\footnote{The continuous version of the outcome was computed based on the number of animals in the household and locality-level prices.} Specifically, the overall and differential attrition rates are 12.2\% and 2.6 percentage points, respectively, and the attrition tests indicate that only a correction for the ATE is needed (see Panel B in Table \ref{tab:empirical_bounds2}). The na\"{i}ve estimate ($\widehat{\Delta}_R$) indicates that the cash transfer increased the probability of ownership of production animals by 5.6 percentage points. This estimate is significant at 1\% and is relative to the mean outcome of 81.5\% for the control group at baseline.

We first report the results of the four corrections: the CiC bounds, the worst-case Manski bounds, the IPW correction, and the Lee bounds. The CiC bounds for the ATE-R are [-0.306, 0.516], and the bounds for the ATE are [-0.357, 0.541]. These bounds are broadly similar to the Manski bounds for the ATE-R and the ATE, which are [-0.418, 0.582] and [-0.489, 0.633]. Thus, neither set of bounds identify the direction of the treatment effect. The IPW point estimates for the ATE-R and  ATE are both effectively equal to the na\"{i}ve treatment effect of 5.6 percentage points and significant at 5\%.  Meanwhile, the Lee bounds for the average treatment effect for the always-responders indicate that the cash transfer increased the probability of ownership of production animals for this group between 3.5 and 6.5 percentage points.\footnote{These bounds are obtained using the equation for binary outcomes in Equation \ref{eq:Lee_binary} of this online appendix.}

This example points to the challenges that researchers face in addressing attrition for binary outcomes, particularly when the object of interest is the ATE. As is the case for the continuous version of this outcome, the test for attrition bias rejects the assumption of internal validity for the population (p=0.000).  This suggests that the na\"{i}ve treatment estimate should not be taken at face value.  Lee bounds, however, are not designed to recover this object, thus the researcher relies only on IPW, CiC, or Manski approaches.   In addition, the IPW correction returns an object that is effectively identical and not statistically different from the na\"{i}ve treatment estimate.\footnote{Although it is true that researchers can add covariates to the IPW (as well as the CiC) corrections, it is notable that conditioning on $Y_0$ has no impact on the IPW-corrected estimate even though it is significantly different across respondents and attritors. }  Thus, the IPW correction is not aligned with the findings from the test of attrition bias.  Finally, CiC and Manski bounds do not identify a sign for the treatment effect and are wide.  Of course, researchers should consider the underlying assumptions of a correction when choosing an approach.\footnote{See Section \ref{sec:appassumptions} in the paper for a discussion of the plausibility of the assumptions for the CiC and the IPW corrections, and  Section \ref{app:empirical_bounds_binary} in this online appendix for a discussion of the Manski and Lee bounds.}

The considerations are similar when evaluating potential corrections for the ATE-R.  In this particular example, the test of attrition bias does not reject (p=0.503), and the IPW correction provides an estimate identical to the na\"{i}ve treatment estimate.  As discussed in Section \ref{sec:appassumptions}, the IPW can return the na\"{i}ve treatment estimate if a correction is not required, even if the IPW assumptions do not hold.  Under monotonicity of response, the Lee approach can bound the treatment effect on the always-responders.  If the percentage of partial compliers is a small subset of the sample, then this object is likely to be similar to the ATE-R.

A final consideration is that, although CiC bounds do not identify a sign for the ATE-R or the ATE in this setting, they may do so in other settings. In line with the discussion in \citet{AI2006}, a potential explanation for the CiC bounds in this setting is the observed time trend in the outcome for the control group. In particular, the average outcome for control respondents decreased by 12 percentage points between baseline and follow-up (see Panel B in Table \ref{tab:empirical_bounds2}).

\begin{table}[htbp!]
\centering
\caption{Ownership of Production Animals} 
\label{tab:empirical_bounds2}
{\small{
\begin{tabular}{lcccccccc}\\
\multicolumn{9}{c}{Panel A. Attrition Corrections} \\
\toprule
Estimator &$\widehat{\Delta}_R$& \multicolumn{1}{c}{{IPW}} & \multicolumn{2}{c}{{CiC Bounds}} & \multicolumn{2}{c}{\parbox{7.5em}{\centering Manski Bounds}} & \multicolumn{2}{c}{{Lee Bounds}}\\
\cmidrule(lr){4-5}\cmidrule(lr){6-7}\cmidrule(lr){8-9}
& & & LB& UB & LB & UB & LB & UB \\
\midrule
& (1) & (2) & (3) & (4) & (5) &(6) & (7) & (8) \\
\midrule
ATT-R  &  & & -0.056&0.762&&&& \\
ATE-R & 0.056***&0.056**&-0.306&0.516&-0.418&0.582&&  \\
ATE  & 0.056***&0.057**&-0.357&0.541&-0.489&0.633&& \\
ATE-Always-responders &&&&&&& 0.035 & 0.065 \\
\bottomrule \\
\end{tabular} \\

\begin{threeparttable}
\begin{tabular}{ccccccccccc}
\multicolumn{11}{c}{Panel B. Attrition Rates, Baseline \& Follow-Up Outcome by Group, and Attrition Tests}\\
    \toprule
\multicolumn{1}{c}{\parbox{5em}{\centering Overall Att. Rate}} & \multicolumn{2}{c}{\parbox{6.5em}{\centering Diff. Att. Rates Test}}& \multicolumn{4}{c}{{\centering Mean Baseline Outcome}} & \multicolumn{2}{c}{\parbox{6.5em}{\centering Mean F/U Outcome}} & \multicolumn{2}{c}{Attrition Tests} \\
 \cmidrule(lr){2-3}
 \cmidrule(lr){4-7}\cmidrule(lr){8-9}\cmidrule(lr){10-11}
   & T-C& $p$-val & CR    & TR    & CA    & TA    & CR & TR & IVal-R & IVal-P \\
    \midrule
12.2&2.6& 0.117& 0.828&0.818&0.699&0.733&0.706&0.762& 0.503 & 0.000 \\
    \bottomrule \\
    \end{tabular}
\begin{tablenotes}[normal, flushleft]
\item \small{ \emph{Notes}: This Table reports the results of the attrition corrections for the binary outcome of value of production animals in the third follow-up of the \emph{Progresa} evaluation. The number of households at baseline and follow-up are 12,299 and 10,799, respectively. Standard errors are bootstrapped. The CiC bounds correspond to those proposed in Section \ref{app:discrete_extension}, the Manski bounds refer to those in \citet{Manski1989}, and the Lee bounds were obtained using the equation in Equation \ref{eq:Lee_binary} of Section \ref{app:other_corrections} in the online appendix. The attrition tests of internal validity for the respondent sub-population (IVal-R) and the study population (IVal-P) correspond to those proposed in \citet{GHO2022}. P-values are reported for these tests. \sym{***} $p<0.01$; \sym{**} $p<0.05$; \sym{*} $p<0.1$.}
\end{tablenotes}
\end{threeparttable}
}}
\end{table}

\section{Simulation Study \label{sec:sim}}
In this section, we examine the finite-sample performance of the CiC attrition corrections.  We also report the results for IPW corrections that rely on unconfoundedness conditional on baseline outcome data, $(Y_{1}(0),Y_{1}(1))\perp R|Y_0$, given their wide use in empirical work as well as their connection to CiC \citep{MM2017, AI2006}.

\subsection{Simulation Design and Estimators}

The simulation design is presented in Panel A in Table \ref{tab:sim_design}.  Treatment is randomly assigned with probability 0.5.  In accordance with Assumption \ref{ass:Mon} in the paper, the potential outcome with and without the treatment are both strictly monotonic in a scalar unobservable, $U_{t}$.  This scalar unobservable has mean zero and consists of a sum of a time-invariant and a time-varying component, $\alpha$ and $\sigma\eta_{t}$, respectively, where the latter is identically distributed across time.\footnote{Via numerical evaluation, we ensure that $U_{t}$ in this design satisfies the time homogeneity assumption conditional on $R(0),R(1)$, which implies Assumption \ref{ass:Unobs}.\ref{TI}.}   Since $U_{t}$ has mean zero, the ATE equals $\beta_1$, which is set to be a quarter of a standard deviation of the potential outcome without the treatment.  The other parameter in $Y_{t}(1)$, $\beta_2$, determines the extent of treatment effect heterogeneity.  

Response in period 1, $R$, is determined by a threshold model in $V$ where the threshold depends on treatment status ($G$) if $a_0\neq a_1$, where $a_0$ and $a_1$ may be interpreted as the cost to response in the control and treatment group, respectively.  The unobservable that determines response, $V$, is a sum of the average unobservable, $\bar{U}$, and idiosyncratic shock, $\eta$.  If $b$ equals zero, then missingness is at random.  The response equation exhibits monotonicity.  Specifically, if we set $a_0>a_1$, then our population consists of always-responders, treatment-only responders and never-responders.  If $a_0=a_1$, then our population consists of always-responders and never-responders only.

\begin{table}[htbp]
	\centering
	\caption{Simulation Design} 
	\footnotesize{\begin{tabular}{ll}
	\multicolumn{2}{l}{Panel A. Data-Generating Process}\\
	\toprule
\multirow{2}{*}{Outcome:} &$Y_{it}(0)=U_{it}$, $Y_{it}(1)=\beta_{1} +\beta_{2}U_{it}+U_{it}$ for $t=0,1$\\
	&where $\beta_{1}=0.25\sigma_{Y_{i1}(0)}$.\\
	[0.5em]
	Treatment:&$G_i\overset{i.i.d.}{\sim} Bernoulli (0.5)$, $D_{i0}=0$, $D_{i1}=G$.\\
	[1em]
	Response: &\parbox{13cm}{$R_i=1\{V_i\geq a_0(1-G_i)+a_1G_i\}$, where $V_i=b\bar{U}_i+\epsilon_i$, $\bar{U}_i=0.5(U_{i0}+U_{i1})$ and $\epsilon_i\overset{i.i.d.}{\sim} N(0,1)$.}\\
	[1em]
	Unobservables:&$\left\{\begin{array}{l}U_{it}=\alpha_i+\sigma\eta_{it}, ~t=0,1; ~(\alpha_i,\eta_{i0},\eta_{i1})\perp \epsilon_i,~\alpha_i\perp (\eta_{i0},\eta_{i1}),\\ \alpha_i\overset{i.i.d.}{\sim} N(0,1)\\
\left(\begin{array}{c}\eta_{i0}\\\eta_{i1}\end{array}\right)\overset{i.i.d.}{\sim} N(0,I_2).\end{array}\right.$\\
	\bottomrule
\\
	\end{tabular}

\begin{tabular}{lccc}
\multicolumn{4}{l}{Panel B. Variants of the Design}\\
\toprule
&Design I& Design II& Design III\\
\midrule
Missing-at-random&No&No&Yes\\
$(U_{0},U_{1})\perp(R(0),R(1))$&$b=1$&$b=1$&$b=0$\\
\midrule
Differential Attrition Rates& Yes&No&Yes\\
&$a_0=\Phi^{-1}\left(\frac{0.3}{\sigma_{V}}\right)$&$a_0=\Phi^{-1}\left(\frac{0.25}{\sigma_{V}}\right)$&$a_0=\Phi^{-1}\left(\frac{0.3}{\sigma_{V}}\right)$\\
&$a_1=\Phi^{-1}\left(\frac{0.2}{\sigma_{V}}\right)$&$a_1=\Phi^{-1}\left(\frac{0.25}{\sigma_{V}}\right)$&$a_1=\Phi^{-1}\left(\frac{0.2}{\sigma_{V}}\right)$\\

\bottomrule
\multicolumn{4}{l}{\parbox{14cm}{\emph{Note}: $\sigma_{Y_1(0)}=\sqrt{Var(Y_{i1}(0))}$, $\sigma_{V}=\sqrt{Var(V)}$, $\Phi^{-1}(q)$ denotes the q$^{th}$ quantile of the standard normal distribution, and $I_2$ denotes the $2 \times 2$ identity matrix.}}
\end{tabular}}
\label{tab:sim_design}
\end{table}

We consider three variants of our design presented in Panel B of Table \ref{tab:sim_design}. All variants feature the same overall attrition rate (25\%), but differ in the values of the parameters that determine response ($a_0$, $a_1$ and $b$).  In Design I, $a_0$ and $a_1$ equal the 30$^{th}$ and 20$^{th}$ percentiles of the distribution of $V$, respectively, whereas we set $b=1$.  The choices of $a_0$ and $a_1$ imply 30\% and 20\% attrition rates in the treatment and control groups, respectively, yielding an overall attrition rate of 25\%.  Setting $b=1$ ensures that $V$ is a function of $\bar{U}$ in addition to $\epsilon$, which violates missing-at-random.  As a result, Design I violates internal validity for the respondents and the study population.  Design II maintains $b=1$ but sets both $a_0$ and $a_1$ to equal the 25$^{th}$ percentile of the distribution of $V$. As a result, it violates internal validity for the study population, while satisfying  internal validity for the respondents since the population consists of always-responders and never-responders in this design.  Finally, in Design III, we set $b=0$, which satisfies missing-at-random. This design maintains $a_0$ and $a_1$ at the same values as in Design I, leading to differential attrition rates.

The primary goal of our simulation analysis is to compare the performance of the CiC attrition corrections and the IPW corrections. For the CiC corrections, we report the simulation results for the four objects we define in Section \ref{sec:id} of the paper: ATT-R, ATU-R, ATE-R, and ATE. For the IPW corrections, we report the simulation results for the ATE-R and the ATE described in Section \ref{sec:ipw} of the paper. The estimator for the ATE-R is given by the difference in mean outcomes among respondents, adjusted by the treatment propensity score conditional on $Y_0$:\footnote{Note that while the general IPW corrections allow the propensity scores to depend on any vector of baseline covariates $X_0$, we use $X_0=Y_0$ to allow direct comparison with the CiC corrections.} 
$$\frac{\sum_{i=1}^{n} Y_{i1} G_i R_i}{\sum_{i=1}^{n} \hat{p}(G_i=1\vert R_i=1,Y_{i0})}-\frac{\sum_{i=1}^{n}Y_{i1} (1-G_i)R_i}{\sum_{i=1}^{n}(1-\hat{p}(G_i=1\vert R_i=1,Y_{i0}))}$$
where $Y_{i1}$ and $Y_{i0}$ are the follow-up and baseline outcomes and $\hat{p}(G_i=1\vert R_i=1,Y_{i0}) = \Phi(\hat{\beta} Y_{i0})R_i$. The estimator for the ATE is given by the difference in mean outcomes among respondents, adjusted by both the treatment and response propensity scores:
$$\frac{\sum_{i=1}^{n}Y_{i1}G_iR_i}{\sum_{i=1}^{n}\hat{p}(G_i=1|R_i=1,Y_{i0})\hat{p}(R_i=1|Y_{i0})}-\frac{\sum_{i=1}^{n}Y_{i1}(1-G_i)R_i}{\sum_{i=1}^{n}(1-\hat{p}(G_i=1|R_i=1,Y_{i0}))\hat{p}(R_i=1|Y_{i0})}$$
where $\hat{p}(R_i=1|Y_{i0}) = \Phi(\hat{\gamma} Y_{i0})$. We estimate  $\hat{\beta}$ and $\hat{\gamma}$ using a one-step GMM procedure. We also trim the sample to guarantee the common support restrictions for the treatment and response scores \citep{Huber2012}.\footnote{We trim observations with a treatment propensity score below 5\% and above 95\% to estimate the ATE-R. To estimate the ATE, we also trim observations with a response propensity score below 5\%.} 

In addition to comparing the performance of the CiC and the IPW corrections, we compare these estimators with the na\"{i}ve difference in group means across respondents at follow-up. This difference is estimated as $\widehat{\Delta}_R\equiv\frac{\sum_{i=1}^nY_{i1}G_iR_i}{\sum_{i=1}^{n}G_iR_i}-\frac{\sum_{i=1}^nY_{i1}(1-G_i)R_i}{\sum_{i=1}^{n}(1-G_i)R_i}$.

\subsection{Simulation Results}

Table \ref{tab:sim_results} reports the simulation results for the CiC and IPW corrections for $n=2,000$ and $\sigma=2$ for constant and heterogeneous treatment effects, $\beta_2=0$ and $\beta_2=1$, respectively. For each of the corrections we examine, we report the mean, bias, standard deviation (SD) and root mean squared error (RMSE).  We also report the true values of the different objects, since they may vary across designs.

\begin{table}[htbp]
  \centering
  \caption{Simulation Results ($n=2,000$, $\sigma=2$)}
  {\footnotesize{
    \begin{tabular}{rclccccclcccc}
    \toprule
          & \multicolumn{6}{c}{Constant Treatment Effects ($\beta_2=0$)}                   & \multicolumn{6}{c}{Heterogeneous Treatment Effects ($\beta_2=1$)} \\
          \cmidrule(lr){2-7}\cmidrule(lr){8-13}
          & True Value  & Estim. & Mean  & Bias  & SD    & RMSE  & True Value & Estim. & Mean  & Bias  & SD    & RMSE \\
                    \cmidrule(lr){2-7}\cmidrule(lr){8-13}
   & \multicolumn{6}{c}{Design I}   & \multicolumn{6}{c}{Design I} \\
             \cmidrule(lr){2-7}\cmidrule(lr){8-13}
          &       & $\widehat{\Delta}_{R}$  & 0.34  & -0.22       & 0.10  &   0.24   &       &  $\widehat{\Delta}_{R}$  & 0.86  & -0.32 & 0.16  & 0.36 \\
    \multicolumn{1}{l}{ATT-R} & 0.56  & CiC   & 0.56  & 0.00  & 0.15  & 0.15  & 1.08  & CiC   & 1.08  & 0.00  & 0.19  & 0.19 \\
    \multicolumn{1}{l}{ATU-R} & 0.56  & CiC   & 0.56  & 0.00  & 0.15  & 0.15  & 1.30  & CiC   & 1.30  & 0.00  & 0.27  & 0.30 \\
    \multicolumn{1}{l}{ATE-R} & 0.56  & CiC   & 0.56  & 0.00  & 0.15  & 0.15  & 1.19  & CiC   & 1.18  & 0.00  & 0.23  & 0.23 \\
          &       & IPW  & 0.34  & -0.21 & 0.10  & 0.24  &       & IPW  & 0.87  & -0.32 & 0.16  & 0.35 \\
    \multicolumn{1}{l}{ATE} & 0.56  & CiC   & 0.53  & -0.03 & 0.13  & 0.13  & 0.56  & CiC   & 0.57  & 0.01  & 0.20  & 0.20 \\
          &       & IPW  & 0.30  & -0.26 & 0.13  & 0.29  &       & IPW  & 0.84  & 0.29  & 0.20  & 0.35 \\
                    \cmidrule(lr){2-7}\cmidrule(lr){8-13}
   & \multicolumn{6}{c}{Design II}   & \multicolumn{6}{c}{Design II} \\
          \cmidrule(lr){2-7}\cmidrule(lr){8-13}
          &       &  $\widehat{\Delta}_{R}$   & 0.56  & 0.00      & 0.11  &  0.11     &       &  $\widehat{\Delta}_{R}$  & 1.20  & 0.00  & 0.17  & 0.17 \\
    \multicolumn{1}{l}{ATT-R} & 0.56  & CiC   & 0.56  & 0.00  & 0.15  & 0.15  & 1.20  & CiC   & 1.19  & 0.00  & 0.20  & 0.20 \\
    \multicolumn{1}{l}{ATU-R} & 0.56  & CiC   & 0.56  & 0.00  & 0.15  & 0.15  & 1.19  & CiC   & 1.19  & 0.00  & 0.27  & 0.20 \\
    \multicolumn{1}{l}{ATE-R} & 0.56  & CiC   & 0.56  & 0.00  & 0.15  & 0.15  & 1.19  & CiC   & 1.19  & 0.00  & 0.23  & 0.23 \\
          &       & IPW  & 0.56  & 0.00  & 0.11  & 0.11  &       & IPW  & 1.20  & 0.00  & 0.17  & 0.17 \\
    \multicolumn{1}{l}{ATE} & 0.56  & CiC   & 0.56  & 0.00  & 0.13  & 0.13  & 0.56  & CiC   & 0.61  & 0.05  & 0.20  & 0.21 \\
          &       & IPW  & 0.56  & 0.00  & 0.13  & 0.13  &       & IPW  & 1.25  & 0.69  & 0.20  & 0.72 \\
                              \cmidrule(lr){2-7}\cmidrule(lr){8-13}
   & \multicolumn{6}{c}{Design III}   & \multicolumn{6}{c}{Design III} \\
                    \cmidrule(lr){2-7}\cmidrule(lr){8-13}
          &       & $\widehat{\Delta}_{R}$  & 0.56  &      0.00 & 0.12  & 0.12      &       &  $\widehat{\Delta}_{R}$   & 0.56  & 0.00  & 0.18  & 0.18 \\
    \multicolumn{1}{l}{ATT-R} & 0.56  & CiC   & 0.56  & 0.00  & 0.15  & 0.15  & 0.56  & CiC   & 0.56  & 0.00  & 0.20  & 0.20 \\
    \multicolumn{1}{l}{ATU-R} & 0.56  & CiC   & 0.56  & 0.00  & 0.15  & 0.15  & 0.56  & CiC   & 0.57  & 0.01  & 0.28  & 0.20 \\
    \multicolumn{1}{l}{ATE-R} & 0.56  & CiC   & 0.56  & 0.00  & 0.15  & 0.15  & 0.56  & CiC   & 0.56  & 0.01  & 0.23  & 0.23 \\
          &       & IPW  & 0.56  & 0.00  & 0.12  & 0.12  &       & IPW  & 0.56  & 0.00  & 0.18  & 0.18 \\
    \multicolumn{1}{l}{ATE} & 0.56  & CiC   & 0.56  & 0.00  & 0.13  & 0.13  & 0.56  & CiC   & 0.57  & 0.01  & 0.19  & 0.20 \\
          &       & IPW  & 0.56  & 0.00  & 0.12  & 0.12  &       & IPW  & 0.56  & 0.00  & 0.18  & 0.18 \\
          \bottomrule
    \end{tabular}%
    }}
    \begin{minipage}{\textwidth}
\scriptsize{\emph{Notes}: The simulation results provided above are based on 1,000 simulation replications.  CiC denotes the Changes-in-Changes estimator of the relevant object. IPW refers to the inverse probability weighting estimator for each object of interest.
$\widehat{\Delta}_R$ represents the difference in group means between treatment and control respondents at follow-up. The bias of this difference in group means is calculated relative to the true ATE-R.}
\end{minipage}
  \label{tab:sim_results}%
\end{table}%

In all three designs we consider, the CiC correction has little or no bias for the relevant object of interest.  The performance of the IPW correction, however, depends on the design in question.  In Design I, where internal validity for the study population is violated, the IPW correction for both the ATE-R and the ATE exhibit a substantial bias regardless of treatment effect heterogeneity.  In Design II, where only internal validity for the respondents holds, the IPW correction for the ATE-R has a negligible bias, but the correction for the ATE is biased when there is treatment effect heterogeneity.   In Design III, there is missing-at-random, and as a result, the IPW corrections exhibit no bias. 

In this simulation design, the IPW correction for the ATE-R tends to have a smaller standard deviation relative to the CiC correction for that same object.  As a result, in Designs II and III, where a correction for the ATE-R is not warranted, the IPW correction for the ATE-R has a slightly smaller RMSE than the CiC.  However, when a correction is warranted as in Design I (II), the CiC correction for the ATE-R and the ATE (ATE-R) has a substantially smaller RMSE relative to IPW, given the latter's sizeable bias.

\end{document}